\documentclass{article} 

\usepackage[margin = 2cm]{geometry}
\usepackage{bm}
\usepackage[countmax]{subfloat}
\usepackage{anyfontsize}%
\usepackage{multirow}
\usepackage{footnote}
\usepackage{url}
\usepackage{amsmath}
\usepackage{mathrsfs}
\usepackage{algorithm}%
\usepackage{algorithmicx}%
\usepackage{algpseudocode}%
\usepackage{listings}%
\usepackage{amssymb}
\usepackage{array}
\usepackage{graphicx}
\usepackage{float}
\usepackage{natbib}
\newcolumntype{P}[1]{>{\centering\let\newline\\\arraybackslash\hspace{0pt}}p{#1}}

\begin{document}

\title{Estimating Methane Emissions from the Upstream Oil and Gas Industry Using a Multi-Stage Framework}

\author{Augustine Wigle \and Audrey B\'{e}liveau}


\maketitle

\abstract{

Measurement-based methane inventories, which involve surveying oil and gas facilities and compiling data to estimate methane emissions, are becoming the gold standard for quantifying emissions. However, there is a current lack of statistical guidance for the design and analysis of such surveys. The only existing method is a Monte Carlo procedure which is difficult to interpret, computationally intensive, and lacks available open-source code for its implementation. We provide an alternative method by framing methane surveys in the context of multi-stage sampling designs. We contribute estimators of the total emissions along with variance estimators which do not require simulation, as well as stratum-level total estimators. We show that the variance contribution from each stage of sampling can be estimated to inform the design of future surveys. We also introduce a more efficient modification of the estimator. Finally, we propose combining the multi-stage approach with a simple Monte Carlo procedure to model measurement error. The resulting methods are interpretable and require minimal computational resources. We apply the methods to aerial survey data of oil and gas facilities in British Columbia, Canada, to estimate the methane emissions in the province. An R package is provided to facilitate the use of the methods.
}

\newcommand{\POD}{\text{POD}}
\newcommand{\Hajek}{H\'{a}jek}
\newcommand{\Unif}{\text{Unif}}

\newcommand{\E}{\text{E}}
\newcommand{\Var}{\text{Var}}
\newcommand{\Cov}{\text{Cov}}

\newcommand{\Ed}{\E_{ptq}}
\newcommand{\Vard}{\Var_{ptq}}

\newcommand{\Eme}{\E_{m}}
\newcommand{\Varme}{\Var_{m}}

\newcommand{\Ymeas}{\dot{Y}}

\newcommand{\stageone}{\text{I}}
\newcommand{\stagetwo}{\text{II}}
\newcommand{\stagethree}{\text{III}}

\newcommand{\UI}{U^{\stageone}}
\newcommand{\UII}{U^{\stagetwo}_p}
\newcommand{\UIII}{U^{\stagethree}_{pt}}

\newcommand{\SI}{S^{\stageone}}
\newcommand{\SII}{S^{\stagetwo}_p}
\newcommand{\SIII}{S^{\stagethree}_{pt}}

\newcommand{\sI}{s^{\stageone}}
\newcommand{\sII}{s^{\stagetwo}_p}
\newcommand{\sIII}{s^{\stagethree}_{pt}}
\newcommand{\sIIstar}{s^{\stagetwo *}_p}

\newcommand{\SIIstar}{S^{\stagetwo *}_p}
\newcommand{\SIIIstar}{S^{\stagethree *}_{pt}}

\newcommand{\D}{D}
\newcommand{\ddp}{d_p}
\newcommand{\Qpt}{Q_{pt}}

\newcommand{\Nhp}{N_{h(p)}}
\newcommand{\nhp}{n_{h(p)}}
\newcommand{\Nhm}{N_{h(l)}}
\newcommand{\nhm}{n_{h(l)}}
\newcommand{\Nh}{N_{h}}
\newcommand{\nh}{n_{h}}

\newcommand{\pip}{\pi_p^\stageone} 
\newcommand{\pim}{\pi_l^\stageone} 
\newcommand{\pipm}{\pi_{pl}^\stageone} 

\newcommand{\pit}{\pi_{t\mid p}^\stagetwo} 
\newcommand{\piu}{\pi_{u\mid p}^\stagetwo} 
\newcommand{\pitu}{\pi_{tu\mid p}^\stagetwo} 

\newcommand{\piq}{\pi_{q \mid pt}^\stagethree} 
\newcommand{\pir}{\pi_{r \mid pt}^\stagethree}
\newcommand{\piqr}{\pi_{qr \mid pt}^\stagethree}

\newcommand{\podq}{\phi_{ptq}}
\newcommand{\podr}{\phi_{ptr}}

\newcommand{\phipt}{\phi_{pt\cdot}} 
\newcommand{\phipu}{\phi_{pu\cdot}} 

\newcommand{\phihpt}{\hat{\phi}_{pt\cdot}} 
\newcommand{\phihir}{\hat{\phi}_{pu\cdot}} 
\newcommand{\podmean}{\mu_{pt\cdot}} 

\newcommand{\sumppop}{\sum_{p \in \UI}} 
\newcommand{\summpop}{\sum_{l \in \UI}} 
\newcommand{\sumtpop}{\sum_{t \in \UII}} 
\newcommand{\sumupop}{\sum_{u \in \UII}} 
\newcommand{\sumqpop}{\sum_{q \in \UIII}} 
\newcommand{\sumrpop}{\sum_{r \in \UIII}}

\newcommand{\sumpsamp}{\sum_{p \in \SI}}
\newcommand{\summsamp}{\sum_{l \in \SI}}
\newcommand{\sumtsamp}{\sum_{t \in \SII}}
\newcommand{\sumusamp}{\sum_{u \in \SII}}
\newcommand{\sumqsamp}{\sum_{q \in \SIII}}
\newcommand{\sumrsamp}{\sum_{r \in \SIII}}

\newcommand{\sumuneq}{\sum_{u \neq t}}

\newcommand{\sumtstarsamp}{\sum_{t \in \SIIstar}}
\newcommand{\sumustarsamp}{\sum_{u \in \SIIstar}}

\newcommand{\Yptq}{Y_{ptq}} 
\newcommand{\Yb}{\overline{Y}}
\newcommand{\Ypt}{\Yb_{pt\cdot}}
\newcommand{\Ypu}{\Yb_{pu\cdot}}
\newcommand{\Yp}{\Yb_{p\cdot\cdot}}
\newcommand{\Ym}{\Yb_{m\cdot\cdot}}

\newcommand{\Yptqmeas}{\Ymeas_{ptq}}

\newcommand{\Yhptq}[1]{\hat{Y}^{\text{#1}}_{ptq}}
\newcommand{\Yhpt}[1]{\hat{\Yb}^{\text{#1}}_{pt\cdot}}
\newcommand{\Yhp}[1]{\hat{\Yb}^{\text{#1}}_{p\cdot\cdot}}

\newcommand{\Yhptl}[1]{\hat{Y}^{\text{#1}}_{ptr}}
\newcommand{\Yhpu}[1]{\hat{\Yb}^{\text{#1}}_{pu\cdot}}
\newcommand{\Yhm}[1]{\hat{\Yb}^{\text{#1}}_{l\cdot\cdot}}

\newcommand{\Th}[1]{\hat{T}^{\text{#1}}{}}

\newcommand{\uptq}{u_{ptq}}
\newcommand{\hptq}{a_{ptq}}
\newcommand{\V}{\text{V}}
\newcommand{\Vh}{\hat{\V}}
\newcommand{\Vtilde}{\tilde{\V}}

\newcommand{\Vtot}[1]{{\V}^{\text{3 stage, #1}}}
\newcommand{\Vone}[1]{{\V}^{\stageone, #1}}
\newcommand{\Vtwo}[1]{{\V}^{\stagetwo, #1}}
\newcommand{\Vthree}[1]{{\V}^{\stagethree, #1}}

\newcommand{\Vhtott}{\hat{\V}^{\text{3 stage}}}
\newcommand{\Vhonee}{\hat{\V}^{\stageone}}
\newcommand{\Vhtwoo}{\hat{\V}^{\stagetwo}}
\newcommand{\Vhthreee}{\hat{\V}^{\stagethree}}

\newcommand{\Vp}{\V_{p\cdot\cdot}}
\newcommand{\Vhp}[1]{\hat{\V}^{\text{#1}}_{p\cdot\cdot}}

\newcommand{\Vpt}{\V_{pt\cdot}}
\newcommand{\Vhpt}[1]{\hat{\V}^{\text{#1}}_{pt\cdot}}

\newcommand{\Varone}{\Vtilde^{\stageone}}
\newcommand{\Vartwo}{\Vtilde^{\stagetwo}}
\newcommand{\Varthree}{\Vtilde^{\stagethree}}
\newcommand{\Varmsmt}{\Vtilde^{m}}
\newcommand{\Vptqm}{\Vtilde^{\text{Total}}}

\newcommand{\tauh}{\tilde{\tau}}

\newcommand{\Ymeasvec}{\bm{\Ymeas}}
\newcommand{\Ytruevec}{\bm{Y}}

\newcommand{\sumb}{\sum_{b = 1}^B}

\newcommand{\sumh}{\sum_{h = 1}^H}
\newcommand{\sumpsampstrata}{\sum_{\substack{p \in \SI, \\ p \in \Omega_{h}}}}
\newcommand{\summsampstrata}{\sum_{\substack{l \in \SI, \\ l \in \Omega_{h}}}}

\newcommand{\sumc}{\sum_{c = 1}^{5000}}

\newcommand{\prodp}{\prod_{p \in \SI}}
\newcommand{\prodtstar}{\prod_{t\in \SIIstar}}
\newcommand{\prodt}{\prod_{t \in \SII}}
\newcommand{\prodtno}{\prod_{t \in \SI\setminus \SIIstar}}

\section{Introduction} 

Significant reductions in methane emissions from key human-related sources are needed to mitigate the worst effects of climate change \citep{lee_ipcc_2023}. An area with important potential for reduction efforts is the upstream oil and gas (O\&G) sector, leading many countries to set targets for methane emissions reductions in  the O\&G industry \citep{climate_and_clean_air_coalition_global_2021,environment_and_climate_change_canada_pan-canadian_2016}. In order to assess progress and create effective reduction strategies, accurate estimates of methane emissions from upstream O\&G are needed, along with accompanying uncertainties. Multiple studies have shown that official estimates of methane emissions in Canada significantly underestimate emissions and highlight the value of measurement-based approaches where active O\&G facilities are surveyed with measurement technology and the resulting data are used to inform an inventory \citep{tyner_where_2021, mackay_methane_2021, johnson_creating_2023, conrad_futility_2023, conrad_measurement-based_2023}. Measurement-based inventories have the potential to provide accurate and data-driven estimates of emissions, as well as insights into which types of facilities should be targeted for reductions efforts. 

In the analysis of survey data to estimate measurable emissions, four sources of uncertainty must be considered. First, it is not practical due to cost and time constraints to measure methane at every facility in the population of interest, such as an entire province or an entire country. This leads to sampling error - that is, the error in the estimate due to not measuring every facility in the population (results will vary if we sample different subsets of facilities). Similarly, it is also impractical to survey facilities every day within a time period of interest, leading to sampling error due to the days. Third, the measurement technology can fail to detect methane at emitting sources even when the emission rate is above its minimum detectable limit, leading to detection uncertainty. Finally, measurement technologies are subject to measurement uncertainty, that is, when methane is detected, the estimated emission rate is not exactly equal to the true emission rate, and is likely biased. These four sources of uncertainty must be considered when creating measurement-based estimates of methane emissions.

Measurement-based data has been collected in multiple surveys in Canada using Bridger Photonics, Inc's Gas-mapping LiDAR (GML) plane-mounted technology \citep{johnson_creating_2023, conrad_futility_2023, conrad_measurement-based_2023}. The GML technology has been studied through extensive controlled releases \citep{johnson_blinded_2021, tyner_where_2021} and both the probability of detection and measurement error of the technology have been characterised \citep{conrad_robust_2023}. These properties correspond to detection uncertainty and measurement uncertainty as described above. The probability of detection function and measurement error distribution can readily be used to quantify these sources of error in methane emissions inventories.

A simulation approach to creating measurement-based inventories from the aerial survey data collected using GML was recently developed and applied in Canada \citep{johnson_creating_2023, conrad_futility_2023, conrad_measurement-based_2023}. To the best of the author's knowledge, it is the only available approach to create measurement-based inventories from aerial survey data in the current literature. The approach separately estimates ``measurable" emissions, which are emissions that are detectable by the GML aerial survey technology, and ``unmeasurable" emissions, which are emissions below the minimum detectable limit of the measurement technology. Only the measurable portion of emissions can be estimated using the GML data, while the unmeasurable portion requires additional methods which rely on auxiliary data from surveys of facilities on the ground using optical gas imaging (OGI). 

The approach to estimating the measurable portion of emissions developed in \citet{johnson_creating_2023} uses a nested Monte Carlo (MC) algorithm to account for measurement error, sampling error, and detection error. First, a MC procedure which accounts for measurement and detection uncertainty (the initial MC) is performed. Next, for every iteration of the initial MC, a mirror match bootstrap procedure \citep{sitter_resampling_1992} is followed in order to scale the sample estimates to the population level and account for sampling error due to the facilities. The algorithm requires computationally intensive methods including numerical integration and root-finding methods to solve multiple equations at each iteration of the initial MC. Additionally, both the initial MC and mirror-match bootstrap are complex and open-source code for their implementation is not available. 

The primary objective of this work is to propose an alternative framework for estimating the measured portion of measurement-based methane inventories and their uncertainties which is easy to interpret and implement. We accomplish this goal by viewing the problem through a statistical survey lens, treating non-detection as survey non-response, and using established multi-stage weighting methods \citep{sarndal_model_1992, wu_sampling_2020}. This provides analytical formulas which account for sampling and detection uncertainty and have desirable statistical properties, such as unbiasedness. Further, we present decompositions of the variance formulas which are useful in identifying important contributors to the uncertainty of the inventory. Additionally, we propose a straightforward three-step MC which incorporates measurement uncertainty. Further, an R package which implements the described analysis was written and is freely available. This alternative method and accompanying R package will make measurement-based inventories more accessible, in accordance with the FAIR principles (Findable, Accessible, Interoperable, Reusable) endorsed by The Intergovernmental Panel on Climate Change (the United Nations body for assessing the science related to climate change) \citep{pirani_implementation_2022}. To the best of the authors' knowledge, this is the first paper addressing the development of measurement-based inventories from a statistical foundation.

The paper is organized as follows: In Section \ref{sec:data}, we provide the necessary background on the motivating dataset and the technology used to collect the data. In Section \ref{sec:methods}, we cast the problem under a design-based statistical framework and explain how weighting methods for multi-stage sampling can be applied in this situation, providing two options for estimators and the necessary modifications to the methodology required for each option. We provide a simple MC procedure that can be combined with the multi-stage framework to incorporate measurement uncertainty into the estimate and its variance.
We apply the methods to the motivating dataset in Section \ref{sec:analysis}, providing insights about key emitters and important sources of uncertainty. We conclude with a discussion of the results and potential future areas of development in Section \ref{sec:disc}.

\section{Aerial Survey Data from British Columbia} \label{sec:data}

Aerial survey data were collected using GML in the province of  British Columbia (BC), Canada in 2021. The survey was the basis for developing a measurement-based methane inventory from the province's O\&G industry \citep{johnson_creating_2023}. To the best of our knowledge, the survey was the first of its kind. The British Columbia Oil and Gas Research and Innovation Society (BC OGRIS) and the Methane Emissions Research Collaborative (MERC) provided anonymized aerial survey data to facilitate the development of new measurement-based methods for estimating methane emissions. Some anonymized data from the BC survey have since been published \citep{johnson_creating_2023}, and subsequent surveys following similar designs to the BC survey have been performed in other provinces in Canada \citep{conrad_futility_2023, conrad_measurement-based_2023}. A subset of the data provided by BC OGRIS and the MERC is available in the R package methaneInventory.

In the BC survey data, the airborne GML technology was used to detect and quantify methane from point sources at selected upstream O\&G facilities. A component is a piece of equipment at an O\&G facility, such as a compressor building or a tank. Measurements made using airborne GML were attributed to specific components \citep{johnson_creating_2023}. A site is a collection of one or more facilities which are close together and were surveyed simultaneously. The plane visited some sites on multiple survey days, resulting in multiple days of measurement data for some components. Some facilities are large and required multiple passes by the plane to cover them completely on each survey occasion. Some components were visible to the plane in multiple passes, resulting in multiple data points for those components on that day. Each time the plane flew over a component is called a pass. In passes where methane was detected, we have available the measured emission rate from GML, the plane's altitude, and the wind speed. The number of passes per day per component ranges between one and five, with a median of two passes per day per component. A total of 543 unique components are identified in the dataset, with a total of 1171 passes with successful detections. 

\begin{figure}
    \centering
    \includegraphics[width = 0.9\linewidth]{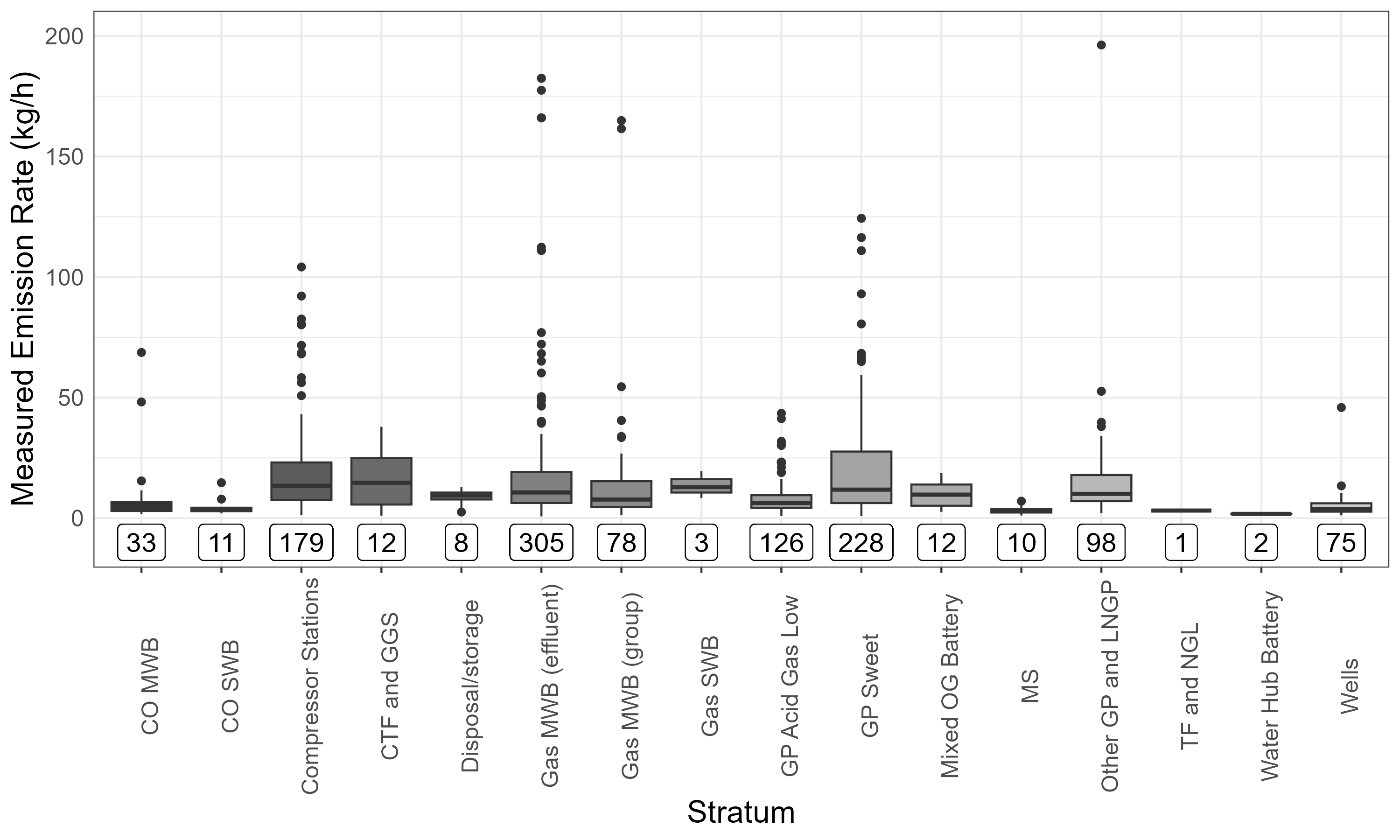}
    \caption{Boxplots of measured emission rates for different facility types (strata). There are two outliers above 200 kg/h in the dataset for the sweet gas plant facility type. Details on the strata and facility types are available in Table \ref{tab:stratum-defs}. The number of non-zero measurements for each stratum is shown in the box directly below each boxplot. Acronyms: GP = gas plant, MWB = multi-well battery, LNGP = liquid natural gas plant, CO = crude oil, CTF = custom treating facility, GGS = gas gathering system, SWB = single-well battery, MS = meter station, TF = tank farm, NGL = natural gas liquids.}
    \label{fig:boxplots}
\end{figure}

Different kinds of facilities are expected to have different levels of emissions, and so it is useful to classify facilities by Petrinex subtype, where Petrinex is Canada's Petroleum Information Network \citep{mackay_methane_2021}. Boxplots of the non-zero measurements for each facility type (stratum) are shown in Figure \ref{fig:boxplots}. Details about the definition of facility types and strata are given in Section \ref{sec:specificdesign} and Table \ref{tab:stratum-defs}. There is wide variation in the distribution of measurements between different strata, in terms of the average, variance, and skewness. There is also variation in the number of non-zero measurements, ranging from one (for tank farms and natural gas liquids) to 305 (for gas multi-well batteries for effluent). There are two strata which are not shown in Figure \ref{fig:boxplots} because they had no non-zero methane emissions measurements (enhanced recovery schemes and water and waste facilities).

The GML's detection abilities are imperfect and sometimes the technology fails to detect emissions. As mentioned in the introduction, work has been done to describe both the probability of detection and the measurement error of the GML technology. A function which calculates the probability of detection (POD) of a given source by the GML  was derived by \citet{conrad_robust_2023} using controlled release data. The POD is a function $\phi$ of the true emission rate ($Y$, in kg/h), altitude of the plane over the source ($a$, in m), and wind speed ($u$, in m/s). The POD of a point source is calculated as
\begin{equation}
    \phi(Y, a, u) = \exp\left[-\left\{\frac{0.244 Y^{1.07}}{\left(\frac{a}{1000}\right)^{2.44}(u+2.14)^{1.69})}\right\}^{-2.53}\right]\label{eq:pod}
\end{equation}
\citep{conrad_robust_2023}. We ignore any error or uncertainty in this function, in line with \citet{johnson_creating_2023}.

In addition, \citet{conrad_robust_2023} derived a probability density function for the true emission rate given a measurement from GML using controlled release data. Letting $\Ymeas$ be the measurement of the true emission rate $Y$, the distribution is
\begin{equation}\label{eq:error-dist}
    Y \mid \Ymeas \sim \text{Log-logistic}(\text{scale}= d\times \alpha \times\Ymeas, \text{shape} = \beta),
\end{equation}
where $d = 0.918$, $\alpha = 0.891$, and $\beta = 3.82$. The mean of $Y$ given $\Ymeas$, by properties of the log-logistic distribution, is
\begin{equation}\label{eq:error-dist-mean}
    d\times\Ymeas \times \frac{\alpha \pi/\beta}{\sin({\pi/\beta})} = d\times \Ymeas = 0.918 \times \Ymeas.
\end{equation}
This means that on average, the true emission rate is equal to $0.918$ times the measured value. This insight provides a straightforward bias correction for measurements.

\begin{table}[H] 
    \centering \caption{Description of facilities, Petrinex codes, and stratum definitions used in the data analysis.} \label{tab:stratum-defs}
    \begin{tabular}{P{0.35\linewidth}|P{0.08\linewidth}|c|P{0.1 \linewidth}|P{0.11\linewidth}} 
        Facility Description & Petrinex Subtype & Stratum Name & Sample size ($\nh$) & Population size ($\Nh$) \\ \hline \hline
        Crude oil single-well battery & 311 & {CO SWB} & 48 & 58 \\ \hline
        Crude oil multi-well group battery & 321 & \multirow{2}{*}{CO MWB} & \multirow{2}{*}{34} & \multirow{2}{*}{40} \\
        Crude oil multi-well proration battery  & 322 & & & \\ \hline
        Gas single-well battery & 351 & Gas SWB & 21 & 28 \\ \hline
        Gas multi-well group battery & 361 & Gas MWB (group) & 54 & 79\\ \hline
        Gas multi-well effluent measurement battery & 362 & Gas MWB (effluent) & 113 & 140 \\ \hline
        Mixed oil and gas battery & 393 & Mixed OG Battery & 16 & 16\\ \hline
        Water hub battery & 395 & Water Hub Battery&  22 & 34\\ \hline
        Gas plant sweet & 401 & GP Sweet & 21 & 25 \\ \hline 
        Gas plant, acid gas flaring ($<1$ t/d sulphur) & 402 & GP Acid Gas Low & 19 & 23 \\ \hline
        Gas plant, acid gas flaring ($>1$ t/d sulphur) & 403 & \multirow{6}{*}{Other GP and LNGP} & \multirow{6}{*}{10} & \multirow{6}{*}{21} \\ 
        Gas plant, acid gas injection & 404 & & &\\ 
        Gas plant, sulphur recovery & 405 & & & \\ 
        Gas plant, fractionation & 407 & & & \\ 
        Liquefied natural gas plant & 451 & & &\\ 
        Gas transporter & 204 & & &  \\  \hline
        Enhanced recovery scheme & 501 & Recovery Scheme & 23 & 27\\ \hline
        Disposal & 503 & \multirow{3}{*}{Disposal and storage} & \multirow{3}{*}{48} & \multirow{3}{*}{77} \\ 
        Acid gas disposal & 504 & & & \\
        Underground gas storage & 505 & & & \\ \hline
        Compressor station & 601 & Compressor station & 45 & 254\\ \hline
        Custom treating facility & 611 & \multirow{2}{*}{CTF and GGS} & \multirow{2}{*}{97} & \multirow{2}{*}{144}\\ 
        Gas gathering system & 621 & & &\\ \hline
        Field receipt meter station & 631 & \multirow{4}{*}{MS} & \multirow{4}{*}{51} & \multirow{4}{*}{91} \\ 
        Interconnect receipt meter station & 632 & & & \\ 
        NEB regulated field receipt meter station & 637 & & & \\ 
        NED regulated interconnect receipt meter station & 638 & & & \\ \hline
        Tank farm loading and unloading terminal & 671 & \multirow{3}{*}{TF and NGL} & \multirow{3}{*}{13} & \multirow{3}{*}{24} \\ 
        Third party tank farm loading and unloading terminal & 673 & & & \\ 
        Natural gas liquids hub terminal & 676 & & & \\ \hline
        Surface waste facility & 701 & \multirow{3}{*}{Water and Waste} & \multirow{3}{*}{13} & \multirow{3}{*}{20} \\ 
        Water source & 901 & &&\\ 
        Water source battery & 902 & && \\ \hline
        Well - Gas & - & \multirow{4}{*}{Wells} & \multirow{4}{*}{1004} & \multirow{4}{*}{9978} \\ 
        Well - oil & - & & & \\ 
        Well - water & - & & &\\ 
        Well - undefined & - & & & \\ 
    \end{tabular}
\end{table}

\newpage

\section{Methods} \label{sec:methods}

In order to reduce the computational complexity required to estimate a measured methane inventory from aerial survey data, we approach the problem from a design-based viewpoint. 
There are four important sources of uncertainty in the BC data which must be considered when estimating total methane emissions. In this section, we show how a multi-stage approach can be leveraged to estimate provincial methane emissions from survey data, accounting for the four sources of uncertainty in measurement-based survey data. In Section \ref{sec:designoverview}, we explain how a multi-stage framework can apply to the survey data. Methods for estimating methane emissions that consider two sources of sampling uncertainty and detection uncertainty are developed in Sections \ref{sec:estimation}, \ref{sec:hajek}, and \ref{sec:specificdesign}. In Section \ref{sec:msmterror}, we provide an algorithm to incorporate measurement uncertainty into the previous methods, allowing the consideration of all four sources of uncertainty in the estimation. A summary of the notation used in this section is given in SI 1.

\subsection{Measurement-Based Surveys As Multi-Stage Sampling} \label{sec:designoverview}

Multi-stage sampling is a type of probability design where sampling is carried out in the following way:
\begin{enumerate}
    \item A sample $\SI$ of primary sampling units (PSUs) are selected from a population $\UI$ according to a probability design.
    \item For every PSU $p$ in $\SI$, a sample of secondary sampling units (SSUs) $\SII$ is drawn from the population $\UII$ according to a probability design.
    \item For every PSU $p$ in $\SI$ and SSU $t$ in $\SII$, a sample of tertiary sampling units (TSUs) $\SIII$ is drawn from the population $\UIII$ according to a probability design,
\end{enumerate}
and so on. Each stage is defined using a probability design which describes how the units at that stage were sampled, given the results of the sampling in the previous stage(s).
In the context of measurement-based aerial surveys of O\&G facilities, a three-stage approach can be employed to describe the data collection process. We introduce stages to represent the selection of components to sample, the collection of data on different days, and the process of detection of methane in each pass.

In order to apply the methods presented here, the first and second order inclusion probabilities for all three stages of sampling must be known. The first order inclusion probability refers to the probability of a sampling unit being selected in a given stage. The second order inclusion probability refers to the probability of any two units both being selected at a given stage. However, data may not have been collected according to a statistical probability design. Practitioners must then work to find a probability design which matches the process by which data were collected as closely as possible. This may require assumptions and approximations which should be kept in mind when interpreting results. 

In the methane data, the first stage of sampling represents the selection of components to survey. Therefore, the first stage population $\UI$ represents the components at all O\&G facilities in the province of BC, and $\SI$ represents the components which were surveyed. To support application of these methods to other measurement-based surveys, we introduce a generic probability design to represent the first stage in Section \ref{sec:estimation}, while also providing results for the specific design assumed for the BC data in Section \ref{sec:specificdesign}. The generic first stage probability design is described by $\pip$, the probability that component $p$ is in $\SI$, and $\pipm$, the probability that component $p$ and $l$ are both in $\SI$.

The second stage of sampling is the selection of the days on which to survey each component selected into $\SI$. For each component $p \in \SI$, we select a sample of days $\SII$ on which to survey $p$. The second stage population $\UII$ represents the entire time period for which we want to estimate emissions. Typically, methane emissions are estimated on a yearly basis, in which $\UII$ is the year for all $p$. Similarly to the first stage, we use a generic second stage design, and later provide specific results for the BC data. The generic second stage design is described by $\pit$, the probability that we survey component $p$ on day $t$ given that $p$ was selected into $\SI$, and $\pitu$, the probability that we survey component $p$ on days $t$ and $u$ given that $p$ is in $\SI$.

Finally, the third stage represents detection of methane in the passes on each day a component was surveyed. While in the first and second stages we use generic probability designs to allow for different modelling of the selection process for components and days, for the third stage we specify a probability design which is broadly applicable to modelling non-detection in measurement-based surveys. We treat non-detection analogously to non-response in a traditional survey context. The set $\SIII$ is the set of passes in which methane was detected from component $p$ on day $t$, while $\UIII$ is the set of all passes made over component $p$ on day $t$, regardless of detection status. This stage is governed by the probability of detection from component $p$ on day $t$ in pass $q$, $\piq$. We leverage Equation \eqref{eq:pod} to calculate the POD for GML, so that in the BC data, $\piq = \phi(\Yptq, \uptq, \hptq) = \podq$, where $\Yptq$ is the true emission rate, $\uptq$ is the wind speed in m/s, and $\hptq$ is the height of the plane during pass $q$ on day $t$ over component $p$. Further, we assume that detection is independent between passes given the variables used to calculate $\podq$, so that the second order inclusion probability $\piqr$ is given by $\podq\podr$. This design corresponds to Poisson sampling \citep{wu_sampling_2020}.

The three stages of sampling represent three important sources of uncertainty in estimating methane emissions from measurement-based survey data: uncertainty due to the facilities and components sampled (stage I), uncertainty due to the days on which sampling takes place (stage II), and detection uncertainty (stage III). The three-stage sampling framework therefore accounts for these three sources of error in the point estimate and its variance estimate.  We refer to estimators which are unbiased with respect to stages I, II, and III as $ptq$-unbiased. The fourth source of uncertainty in the survey data, measurement uncertainty, must be incorporated outside of the multi-stage framework; see Section \ref{sec:msmterror}.

\subsection{Estimating Methane Emissions Using Multi-Stage Sampling}\label{sec:estimation}

Our target of estimation is the total methane emission rate from all O\&G facilities in BC. We formulate the estimand as
\begin{equation}
    T = \sumppop \Yp = \sumppop \frac{\sumtpop \Ypt}{\D} = \sumppop \frac{\sumtpop \sumqpop \Yptq/\Qpt}{\D}, \label{eq:estimand}
\end{equation}
where $\Qpt$ is the total number of passes over component $p$ on day $t$, and $\D$ is the total number of days in the period of interest. Further, $\Ypt$ and $\Yp$ are the emission rate from component $p$ on day $t$ averaged over the passes, and the emission rate from component $p$ averaged over the days and the passes. $T$ is an adaptation of the commonly used multi-stage total estimator with no averaging, $ \sumppop \sumtpop \sumqpop \Yptq$ \citep{sarndal_model_1992}. In this section, we adapt existing results to accommodate $T$.

An (approximately) $ptq$-unbiased estimator for $T$ is
\begin{equation}
    \Th{} = \sumpsamp \frac{1}{\pip} \Yhp{} = \sumpsamp \frac{1}{\pip} \frac{\sumtsamp \Yhpt{}/\pit}{\D} \label{eq:Th}
\end{equation}
\citep{sarndal_model_1992}, where $\Yhp{}$ is an estimator for the true emission rate from component $p$, averaged over all days and passes, and $\Yhpt{}$ is an (approximately) unbiased estimator for the mean emission rate from component $p$ on day $t$. Possible forms for $\Yhpt{}$ are described later in this subsection and in Section \ref{sec:hajek}.

A variance estimator for \eqref{eq:Th} which accounts for the uncertainty due to the three stages (component sampling, day sampling, and detection uncertainty) is
\begin{equation}
    \Vhtott = \sumpsamp\summsamp \frac{\left(\pipm - \pip \pim\right)}{\pipm}\frac{\Yhp{}}{\pip}\frac{\Yhm{}}{\pim} +\sumpsamp \frac{\Vhp{}}{\pip}, \label{eq:Vhtott}
\end{equation}
\citep{sarndal_model_1992} where
\begin{equation}
    \Vhp{} = \frac{1}{\D^2}\left[\sumtsamp\sumusamp \frac{(\pitu - \pit\piu)}{\pitu}\frac{\Yhpt{}}{\pit}\frac{\Yhpu{}}{\piu} + \sumtsamp \frac{\Vhpt{}}{\pit}\right]
\end{equation}
is an unbiased estimator of the variance of $\Yhp{}$, and $\Vhpt{}$ is an unbiased estimator for $\Yhpt{}$. The true variance is given in SI 1. A 95\% Wald CI for the true total emissions is given by
\begin{equation}
    \Th{} \pm 1.96\times \sqrt{\Vhtott}.\label{eq:wald-ci}
\end{equation}

The variance can be broken down into contributions from each stage, corresponding to three of the four sources of uncertainty in the data. The estimated contribution due to detection uncertainty (stage III), sampling of days (stage II), and sampling of components (stage I) are given by 
\begin{equation}
    \Vhthreee = \max\left\{0, \sumpsamp \frac{1}{(\pip)^2}\frac{1}{\D}\sumtsamp \frac{\Vhpt{}}{(\pit)^2}\right\},  \label{eq:Vh3}
\end{equation}
\begin{equation}
    \Vhtwoo =\max\left\{0, \sumpsamp \frac{\Vhp{}}{(\pip)^2} - \Vhthreee\right\}, \label{eq:Vh2}
\end{equation}
and
\begin{equation}
    \Vhonee = \max\left\{0,\Vhtott - \Vhtwoo - \Vhthreee\right\}, \label{eq:Vh1}
\end{equation}
respectively, where the $\max$ operator avoids negative variance estimates which can arise especially with small sample sizes at stage II and III \citep{sarndal_model_1992}.

Now, a method for estimating $\Ypt$ in the presence of detection error must be chosen. We can use inverse probability weighting (IPW) to account for detection uncertainty and obtain an unbiased estimate of $\Ypt$. As discussed in Section \ref{sec:designoverview}, detection or non-detection in each pass is modelled as Poisson sampling, with success probability $\podq$ for component $p$ on day $t$ in pass $q$. Under this design, the IPW estimate of $\Ypt$ is
\begin{equation}
    \Yhpt{IPW} = \frac{\sumqsamp \Yptq/\podq}{\Qpt} \label{eq:ipw}
\end{equation}
with corresponding variance estimate
\begin{equation}
   \Vhpt{IPW} = \frac{1}{\Qpt^2}\sumqsamp \frac{(1-\podq)}{\podq^2}\Yptq^2 \label{eq:ipw-var-est}
\end{equation}
\citep{wu_sampling_2020}. If the $\podq$ are perfectly known, then the estimator \eqref{eq:ipw} and its variance estimator are $ptq$-unbiased. 

\subsection{Increasing Efficiency Using the \Hajek\ Estimator} \label{sec:hajek}

A drawback to the use of the IPW estimator to correct for detection uncertainty is that it can result in paradoxical estimates and have a large variance in some situations \citep{horvitz_generalization_1952, sarndal_model_1992, wu_sampling_2020}. An alternative to the use of the IPW estimator in \eqref{eq:Th} is the \Hajek\ estimator, which has the form
\begin{equation*}
    \frac{\sumqsamp\Yptq/\podq}{\sumqsamp 1/\podq}
\end{equation*}
\citep{hajek_discussion_1971,sarndal_model_1992}. The \Hajek\ estimator is only approximately $ptq$-unbiased, but has a much smaller variance than the IPW estimator in a variety of cases and is often preferred \citep{sarndal_model_1992, wu_sampling_2020}.
In particular, the \Hajek\ estimator has been shown to have smaller variance in cases where 1) the $\Yptq$ are constant and 2) the sample size is variable. In the context of the motivating dataset, for most components, the emission rates are relatively homogeneous between passes in the same day, approximately satisfying case 1). Due to missed detections, we also have a variable stage III sample size, satisfying case 2). The \Hajek\ estimator therefore has the potential to reduce the variance contribution from detection uncertainty.

However, caution is needed when attempting to use the \Hajek\ estimator in the context of estimating methane emissions. In typical sampling designs, samples of size zero are not allowed, however, due to the low number of passes per day and variation in detection probability, the size of the stage III sample $\SIII$ may be zero for some components and days. When the size of $\SIII$ is zero, the \Hajek\ estimator is undefined ($0/0$). Therefore, in order to make use of the \Hajek\ estimator in estimating methane inventories, one must adjust the stage II sampling design to ensure it is well-defined.

To ensure $\SIII$ is non-empty for use with the \Hajek\ estimator, we define $\SIIstar$ as the set of days where component $p$ is surveyed AND methane is detected at least once, which ensures that all stage III samples are non-empty. This approach is equivalent to the previously described approach if the IPW estimator is used, which we show in SI 1. It results in  modifications to the stage II and III probability designs. Stage I remains unchanged.  

Detection is marginally independent between passes of the plane, so the probability of having any detections for component $p$ on day $t$ given that $p \in \SI$ and $t \in \SII$ is given by
\begin{equation*}
    \phipt = 1- \prod_{q \in \UIII} (1- \podq).
\end{equation*}

The first order inclusion probabilities for the adjusted stage II sampling design are given by
\begin{equation}
    {\pit}^* = \phipt\pit,
\end{equation}
the probability that day $t$ is in $\SIIstar$ given $p \in \SI$. The probability that $t$ and $u$ are both in $\SIIstar$ is
\begin{equation}
    {\pitu}^* = \begin{cases}
        \phipt\phipu\pitu & \text{if } t \neq u \\
        {\pit}^* & \text{if } t = u.
    \end{cases}
\end{equation}

The stage III inclusion probabilities also change as a consequence of the stage II adjustment. In the original sampling design, the inclusion probability for pass $q$ is equal to $\podq$, the probability that methane is detected in pass $q$ given that $p \in \SI$ and $t \in \SII$. In the modified sampling design, the inclusion probability for pass $q$ is the probability that $q \in \SIII$ given that $p \in \SI$ and $t \in \SIIstar$, that is, the probability of detecting methane in pass $q$ given that $p$ is surveyed on day $t$ AND that methane is detected in at least one pass on that day. This additional condition in the modified design means the responses are no longer independent between passes, and so the stage III sampling scheme is no longer Poisson. Instead, the inclusion probabilities are given by
\begin{equation}
    {\piq}^* = \podq/\phipt,
\end{equation}
and the passes are no longer independent, with 
\begin{equation}
    {\piqr}^* = \podq\podr/\phipt
\end{equation}
when $q\neq r$.

Let
\begin{equation}
    \Yhpt{\Hajek} =\frac{\sumqsamp \Yptq\phipt/\podq}{\sumqsamp \phipt/\podq} = \frac{\sumqsamp \Yptq/\podq}{\sumqsamp 1/\podq},
\end{equation}
which is always defined for $t \in \SIIstar$. In the methane data, an approximately unbiased variance estimate is
\begin{equation}
    \Vhpt{\Hajek} = \frac{\phipt}{\Qpt^2}\left[\sumqsamp (1-\podq) \left(\frac{\Yptq - \Yhpt{\Hajek}}{\podq}\right)^2 + (\phipt-1) \left( \sumqsamp \frac{\Yptq- \Yhpt{\Hajek}}{\podq}\right)^2\right].
\end{equation}

Note that $\phipt$ can only be calculated for components on days in which methane was detected in all $\Qpt$ passes, since we require $\Yptq$ to calculate $\podq$. In practice, many components have some passes where detections were missed. For components on days with a mix of detections and non-detections, we use mean imputation (MI) to estimate $\phipt$. That is, in passes with missed detections, we impute the mean of the detection probabilities for the successfully detected passes on that day. Let $\podmean = \sumqsamp \podq / |\SIII |$ be the mean of the detection probabilities among the detected passes, where $|\cdot|$ denotes the number of elements in $\cdot$. Then, let
\begin{equation}
    \phihpt = 1- (1-\podmean)^{|\UIII \setminus \SIII|} \prod_{k \in \SIII}(1-\podq). \label{eq:phiest}
\end{equation}
In practice, we use this estimate in place of the unknown true value $\phipt$ in all estimators presented here.

To estimate the total methane emissions using the \Hajek\ estimator for the component's daily mean emission rates, replace $\SII$ with $\SIIstar$, $\pit$, $\piu$, and $\pitu$ with ${\pit}^*$, ${\piu}^*$, and ${\pitu}^*$, $\Yhpt{IPW}$ with $\Yhpt{\Hajek}$, and $\Vhpt{IPW}$ with $\Vhpt{\Hajek}$ in the methods described in Section \ref{sec:estimation}.

\subsection{First and Second Stage Designs for BC Data}\label{sec:specificdesign}

In this section, we explain the particular stage I and II designs we use to analyse the BC survey data. 

\subsubsection{Stage I: Stratified Cluster Sampling}

Sampling of components was done by first identifying all the active O\&G facilities in the province \citep{johnson_creating_2023}. Then, in an effort to maximise sample size within a constrained budget and to accommodate on-site follow-up surveys needed to estimate the unmeasured inventory, all facilities south of the 58 degree north latitude line were included in the survey because they are more densely populated and easier to access than more remote facilities. All components at facilities included in the survey were sampled at least once, that is, surveyed at least once with one or more passes per survey.

In the analysis, we take a post-stratified approach to stage I. That is, we treat the sample of facilities as if it were an independent SRS of facilities in each stratum and the sample size in each stratum was fixed \citep{smith_post-stratification_1991}. We stratify based on the type of facility, since the emissions distribution differs by facility type, as seen in the data in Figure \ref{fig:boxplots} and in previous studies, e.g. \citet{mackay_methane_2021}.

More specifically, strata were defined based on Petrinex facility subtypes. Some subtypes were merged to form larger strata based on expert knowledge and to improve sample sizes as described in \citet{johnson_creating_2023}. This scheme still leaves some strata with very small sample sizes, as low as one, posing a problem for variance estimation. Guidance has been presented in the statistical literature on the desirable minimum stratum sample size in post-stratification, ranging from 10 to 20 \citep{sarndal_model_1992, cochran_sampling_1977, kish_survey_1965}. In particular, \citet{kish_survey_1965} recommends a minimum sample size of 10 when the true stratum weights are known, such as in our application. \citet{westfall_post-stratified_2011} investigated the minimum sample size needed for accurate estimation in the context of forest inventories via simulation studies and find that stratum sizes of 10 or more gave stable mean estimates. Thus, we merge more strata than the initial analysis to ensure a minimum sample size of 10 per stratum. Subtypes which had a sample size less than 10 were merged with strata with larger sample sizes that had similar characteristics. A remaining subtype with a small sample size which is not similar to any other subtypes (701 - surface waste battery) was merged with other strata which had small sample sizes even after merging with similar strata (901 - water source and 902 - water source battery) to form a stratum of sufficient size. THis results in a total of $H = 18$ strata. Stratum definitions are given in Table \ref{tab:stratum-defs}. Sample and population sizes were derived from data provided by the MERC, and differ from those used in the other published analysis of these data \citep{johnson_creating_2023}. Details on the sample and population sizes are given in Appendix \ref{sec:samp-pop-sizes}. Note that due to differences in how data for wells was reported, they require slightly different methods, see Appendix \ref{sec:wells}.

Since a facility being surveyed means that all components at that facility are surveyed, the selection of components can be viewed as cluster sampling, where all elements of the cluster are surveyed. Let $h(p)$ represent the stratum that component $p$ belongs to, and let $\Omega_h$ be the set of components belonging to stratum $h$. Let $\nhp$ be the number of facilities in stratum $h(p)$ in the stage I sample and $\Nhp$ be the number of facilities in stratum $h(p)$ in the population. Further, let $f(p)$ represent the facility containing component $p$. Then the inclusion probabilities for stage I are given by
\begin{equation}
    \pip = \frac{\nhp}{\Nhp} \label{eq:stage1-probs}
\end{equation}
and
\begin{equation}
    \pipm = \begin{cases}
        \frac{\nhp}{\Nhp} & \text{if } f(p) = f(l) \\
        \frac{\nhp (\nhp-1)}{\Nhp (\Nhp-1)} & \text{if } f(p) \neq f(l), h(p) = h(l) \\
        \frac{\nhp}{\Nhp}\frac{\nhm}{\Nhm} & \text{if } h(p) \neq h(l).\label{eq:stage1-probs2}
    \end{cases}
\end{equation}

Note that the assumed design implies that components in separate strata are independent. Because of this, we can easily calculate stratum-specific total estimates using
\begin{equation*}
    \Th{}(h) = \sumpsampstrata \frac{1}{\pip} \Yhp{},
\end{equation*}
which has corresponding variance estimate
\begin{equation*}
    \Vhtott(h) = \sumpsampstrata\summsampstrata \frac{\left(\pipm - \pip \pim\right)}{\pipm}\frac{\Yhp{}}{\pip}\frac{\Yhm{}}{\pim} +\sumpsampstrata \frac{\Vhp{}}{\pip}.
\end{equation*}
We also have
\begin{equation}
    \Vhtott = \sumh \Vhtott(h),
\end{equation}
which means $\Vhtott(h)$ is an estimate of stratum $h$'s contribution to the variance in the province-wide methane emissions estimate.

\subsubsection{Stage II: Simple Random Sample}

The stage II design describes which days are chosen to survey the components selected in stage I. We will let $\ddp$ be the number of days on which component $p$ was surveyed, whereas $\D$ is the number of days in the population of interest. For example, if the data are being used to compile a yearly inventory, the population is $\D = 365$ days in the year. We assume here that the days where sampling occurred represent a simple random sample from the population of $\D$ days. This is a simplification of the true underlying process, and the implications of this are discussed in Section \ref{sec:disc}. 

For a simple random sample of $\ddp$ out of $\D$ days, the inclusion probabilities are given by
\begin{equation}
    \pit = \frac{\ddp}{\D}
\end{equation}
and 
\begin{equation}
    \pitu= \begin{cases}
        \pit & \text{if } t= u; \\
        \frac{\ddp(\ddp-1)}{\D(\D-1)} & \text{otherwise}.
    \end{cases}
\end{equation}

The general formula for the variance estimate of $\Yhp{}$ can be computationally intensive, so we derived a computationally simpler form specific to the SRS design at stage II when the IPW estimator is used:
\begin{equation*}
    \Vhp{IPW} = \frac{1}{\D^2}\left[\sumtsamp\frac{(\D-\ddp)\D}{\ddp(\ddp-1)}(\Yhpt{IPW})^2 + \frac{\D(\ddp-\D)}{\ddp^2(\ddp-1)}\left(\sumtsamp\Yhpt{IPW}\right)^2 + \sumtsamp\frac{\D}{\ddp}\Vhpt{IPW}\right].
\end{equation*}
An analogous variance formula for the \Hajek\ estimator is
\begin{equation*}
    \Vhp{\Hajek} = \frac{1}{\D^2}\left[\sumtstarsamp \frac{\D\left\{\D-1-\phihpt(\ddp-1)\right\}}{\ddp(\ddp-1)}\left(\frac{\Yhpt{\Hajek}}{\phihpt}\right)^2 + \frac{\D(\ddp - \D)}{\ddp^2(\ddp-1)}\left(\sumtstarsamp \frac{\Yhpt{\Hajek}}{\phihpt}\right)^2 + \sumtstarsamp \frac{\D}{\ddp \phihpt} \Vhpt{\Hajek}\right].
\end{equation*}
Derivations are available in SI 1.

An estimation issue arises for components where $\ddp = 1$ when using IPW, and components for which there is only one day with detections when using \Hajek. The variance estimation formulas result in undefined expressions under these conditions. If no methane was detected at component $p$ on any survey days, we conservatively assume that this component has an emission rate of zero for the population of $\D$ days and therefore contributes zero to the total and zero to the variance. If component $p$ has only one day of measurement data (either due to $\ddp =1$ for IPW or by only having detections on one day for \Hajek), we pool the $\Vhp{}$ from all other components in the same stratum which were surveyed more than once and impute the average. There are 17 components that were surveyed only once ($\ddp = 1$), and 208 components that have only one day of non-zero measurement data.

\subsection{Estimation Accounting for Measurement Error} \label{sec:msmterror}

In the previous methods, in line with survey sampling literature, we treat the values of $\Yptq$ as fixed and known. In reality, the measurements vary around the true emission rate. To accommodate this, we introduce a superpopulation model by assuming the true emission rate $\Yptq$ follows a distribution given the measured emission rate $\Yptqmeas$. In particular, we can assume that the true emission rate follows the distribution in Equation \eqref{eq:error-dist}. We denote this measurement model as $m$. Let $\Ed$ and $\Vard$ denote the expectation and variance with respect to the $ptq$-design, that is, stages I, II, and III of the sampling design, treating the measured emission rates as fixed and known. Now, let $\Eme$ and $\Varme$ denote the expectation and variance with respect to the measurement model $m$ while the sampling design is fixed. Finally, let $\E$ and $\Var$ represent the total expectation and variance, that is, the expectation and variance with respect to both the $ptq$-design and measurement model $m$. Previously, our target of estimation was $T = \sumppop \Yp$, which is the expectation of $\Th{}$ with respect to the $ptq$-design. Now, we introduce the total expectation of $\Th{}$,
\begin{equation}
    \tau = \Eme(T) = \Eme[\Ed(\Th{})] = \E(\Th{}),
\end{equation}
to be our new target of estimation. 
Given a sample of size $B$ drawn from the distribution of $\Yptq$ given $\Yptqmeas$, we can calculate $B$ estimates of $T$, $\Th{}{}^b$, $b = 1$, $\dots$, $B$. Then we estimate $\tau$ with
\begin{equation}
    \tauh = \frac{\sumb \Th{}{}^b}{B},
\end{equation}
where the hat overscore denotes an estimator with respect to the $ptq$-design and the tilde overscore denotes an estimator with respect to both the $ptq$-design and model $m$.

The total variance of $\tauh$, $\Var(\tauh)$, can be estimated according to
\begin{equation} \label{eq:tauhvar}
    \Vptqm = \Varmsmt + \Varone + \Vartwo + \Varthree,
\end{equation}
where 
\begin{equation*}
    \Varmsmt = \frac{\sumb (\Th{}{}^b - \tauh)^2}{B-1}
\end{equation*}
is the estimated contribution to the total variance of $\tauh$ from measurement uncertainty, and
\begin{equation*}
    \tilde{\V}^{j} = \frac{\sumb \hat{\V}^{j}{}^b}{B}
\end{equation*}
for $j = \stageone, \stagetwo,\stagethree$ represents the estimated contribution due to stages I, II, and III of the $ptq$-design respectively. A derivation of Equation \eqref{eq:tauhvar} is available in SI 1.

An algorithm to estimate the methane emissions total and its variance with respect to the $ptq$-design and measurement model $m$ is given in Algorithm \ref{alg:mealg}. This algorithm addresses the four key sources of uncertainty in the BC data: sampling uncertainty from components and days, detection uncertainty, and measurement uncertainty.
\begin{algorithm}
\caption{Monte Carlo procedure to account for measurement error.}
\label{alg:mealg}
    \begin{algorithmic}
        \For{$b$ in $1, \dots, B$}
            \For{$p \in \SI$}
                \For{$t \in \SII$}
                    \For{$q \in \SIII$}
                        \State Draw true emission rate $\Yptq^b$ from the distribution of true emission rate given measurement $\Yptqmeas$ using the distribution in \eqref{eq:error-dist}
                        \State Calculate $\podq^b = \POD(\Yptq^b, \uptq, \hptq)$ using Equation \eqref{eq:pod}
                    \EndFor
                    \State If \Hajek\ estimator used, calculate $\phihpt^b$
                \EndFor
            \EndFor
            \State Calculate $\Th{}{}^b$, $\Vhonee{}^b$, $\Vhtwoo{}^b$, and $\Vhthreee{}^b$ using equations \eqref{eq:Th}, \eqref{eq:Vh1}, \eqref{eq:Vh2}, and \eqref{eq:Vh3}.
        \EndFor
        \State Calculate $\tauh$, $\Vptqm$, and CI for $\tauh$.
    \end{algorithmic}
\end{algorithm}

\section{Data Analysis} \label{sec:analysis}

In this section, we apply the methods described in Section \ref{sec:methods} to calculate a provincial methane inventory for the data described in Section \ref{sec:data}. The data analysis was performed with the R package \texttt{methaneInventory}, available on GitHub from \url{https://github.com/augustinewigle/methaneInventory}. A tutorial on the use of this package is available in SI 2.

We analyse the BC data in eight different ways, using different approaches to estimate $\Ypt$ (either IPW or \Hajek\ estimator), different approaches to model the selection of days in stage II, and different approaches to address measurement uncertainty. We compare the results to the published inventory compiled in \citet{johnson_creating_2023} using the nested MC approach.  The different stage II and measurement uncertainty approaches are detailed below.

We use two different values for $\D$ in the stage II design to illustrate the ability of our method to incorporate uncertainty due to the number of days sampled. 
First, we assume that the population of interest is represented by the $\ddp$ days on which we sampled component $p$. In other words, we let $\D = \ddp$ for component $p$. From this perspective, the sampling at stage II contributes nothing to the variance, since we are sampling the entire population. This is an equivalent treatment to the main analysis presented by \citet{johnson_creating_2023}. Second, we set $\D = 365$, treating the year as the population of interest. Changing the $\D$ value only impacts the variance estimate and does not change the point estimate.

We also consider measurement error in two different ways. First, we investigate the possibility of avoiding any MC steps by performing a bias correction to the measured emission rates. We multiply all measured emission rate values by the bias correction factor in Equation \ref{eq:error-dist-mean}, and then apply the methods in Section \ref{sec:estimation} without using the measurement uncertainty MC. This provides a rough estimate where we account for measurement bias, but assume that measurement uncertainty is otherwise negligible.  The other approach is to consider both the bias and uncertainty coming from measurement error by applying Algorithm \ref{alg:mealg}. No bias correction to the measured rates is required with this approach because the MC simultaneously corrects for bias and variance. When the measurement uncertainty algorithm is applied, we use $B = 8000$ MC iterations. Convergence was assessed by plotting $\Vtilde^{\text{I}} + \Vtilde^{\text{II}} + \Vtilde^{\text{III}}$ versus the number of MC iterations for each stratum. The different approaches to measurement error can impact  both the point estimate and the variance estimate.

\begin{figure}
    \centering
    \includegraphics[width = 0.9\linewidth]{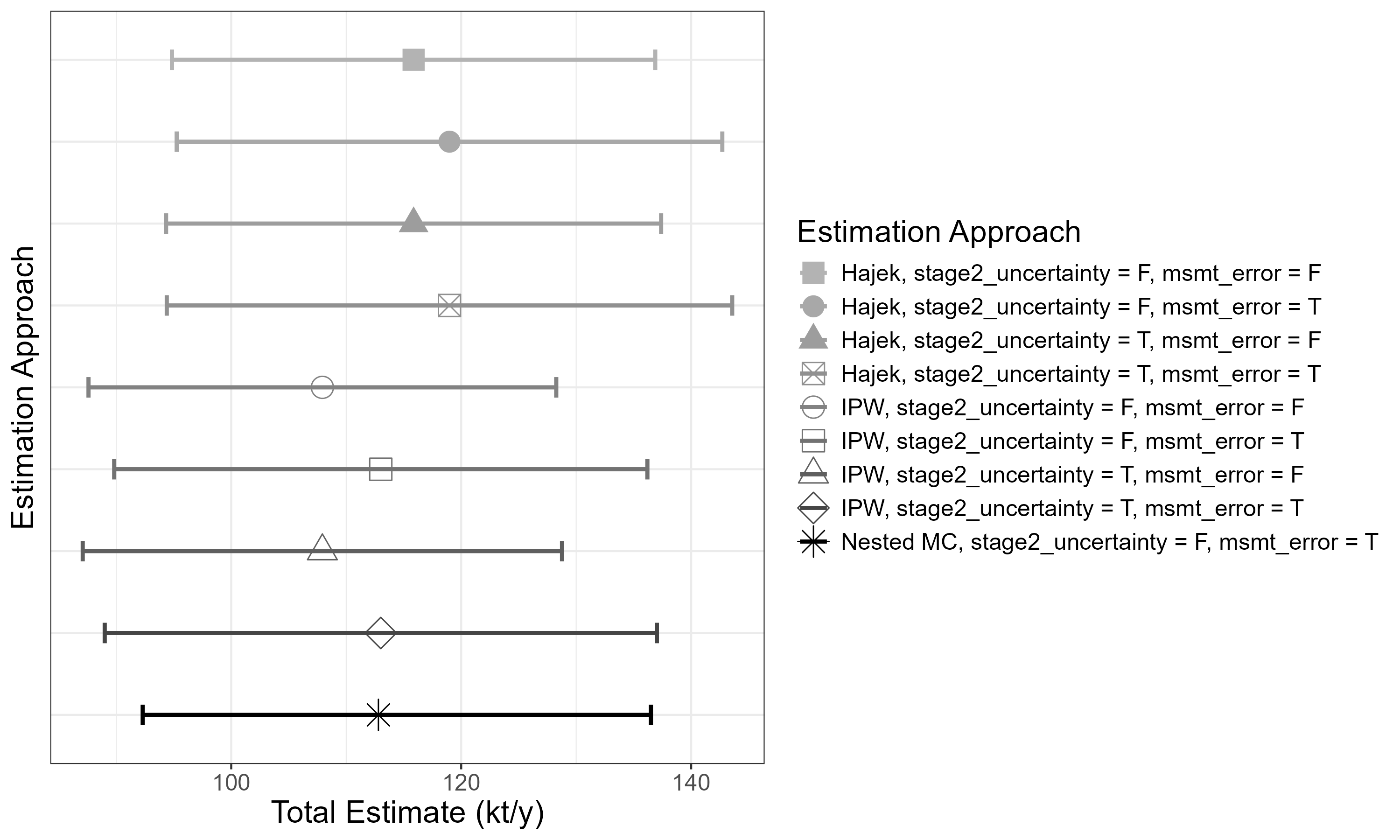}
    \caption{Point estimates and 95\% CIs for all methods compared on the BC data. In the legend, stage2\_uncertainty = F indicates that $D=\ddp$, whereas stage2\_uncertainty = T indicates that $\D = 365$. Similarly, msmt\_error = F means the bias-correction method without the algorithm was used while msmt\_error = T means the measurement uncertainty MC algorithm was used.}
    \label{fig:allmethods}
\end{figure}

Estimates of the total measurable methane emissions for the province of BC in 2021 for eight different multi-stage approaches and the nested MC approach of \citet{johnson_creating_2023} are shown in Figure \ref{fig:allmethods}. In particular, the results for the nested MC approach (represented by the black asterisk in Figure \ref{fig:allmethods}) are very similar to that of the results using the multi-stage approach with IPW where stage II uncertainty is ignored and measurement uncertainty is considered (represented with a dark grey open square in Figure \ref{fig:allmethods}). These approaches consider the same sources of uncertainty. There are small differences in the CI lengths which may be due to the normality assumption relied on in the multi-stage approach which is not used in the nested MC. 

The largest methane emissions estimate is 119.0 kt/y, obtained using the multi-stage \Hajek\ approach with measurement uncertainty, while the lowest estimate is 107.9 kt/y which is produced by the multi-stage approach with IPW and ignoring measurement uncertainty.  The approach with the narrowest 95\% CI is the multi-stage IPW approach which ignores stage II and measurement uncertainty while the approach with the widest 95\% CI is the multi-stage \Hajek\ approach which considers both stage II and measurement uncertainty. In general, using the \Hajek\ estimator over IPW gives larger point estimates and variance estimates. This is likely due to a slight upwards bias observed when using the \Hajek\ estimator in a simulation study which mimicked the BC data as closely as possible (see SI 1). Additionally, as expected, considering additional sources of uncertainty (stage II, measurement) increases the variance estimate. Considering measurement error with the MC algorithm not only increases the variance estimate, but also increases the point estimate. This suggests that multiplying measurements by a bias correction factor may not be sufficient to capture the effects of measurement uncertainty on the point estimate. The distribution of true emission rates given measurements is right-skewed, which the bias-correction factor approach may be unable to capture. To give a full picture of the uncertainty, and to take advantage of unbiasedness properties of IPW, the remaining results presented in this section are from the multi-stage approach with IPW considering both stage II and measurement uncertainty, unless otherwise specified.

The total estimates, variance estimates by source, and total variance estimate for each stratum is shown in Table \ref{tab:stratumsummaries}. There is important variation between strata. Two strata had no non-zero methane measurements in the survey data, and so we conservatively estimate that they emit zero methane over the year (enhanced recovery schemes, water and waste), which are not shown. Beside those two strata, the lowest emitting stratum is tank farms and natural gas liquid storage, which is estimated to emit 0.03 kt/y. This contrasts with the highest emitting stratum, compressor stations, which is estimated to emit 45.49 kt/y.
The top four highest emitting strata account for over 80\% of the total estimated emissions for BC. Compressor stations accounted for an estimated 40.3\% of total provincial emissions. The next highest-emitting strata are sweet gas plants, wells, and multi-well gas batteries (effluent), contributing 15.2, 13.1, and 13.0\% of emissions respectively. The remaining strata account for an estimated 18.4\% of the total emissions. These findings align well with those obtained using the nested MC method, for example, they find the same top four emitting strata.

\begin{figure}
    \centering
    \includegraphics[width = 0.7\linewidth]{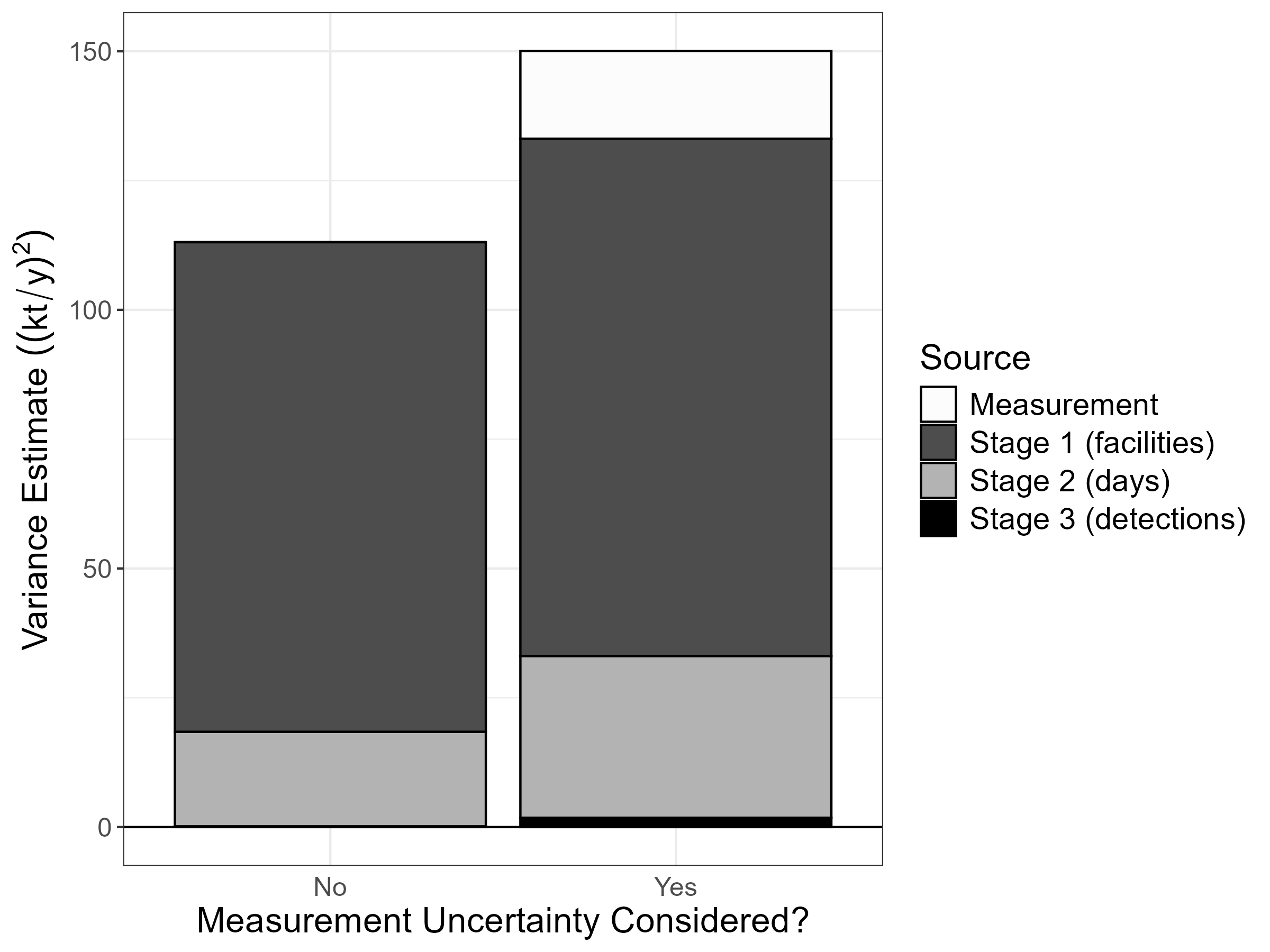}
    \caption{The estimated variance contributions for the provincial total from different uncertainty sources, with and without considering measurement error through the MC algorithm. Results are from the IPW, stage 2 uncertainty, measurement uncertainty model.}
    \label{fig:varbysource}
\end{figure}

Turning to the variance decomposition, the variance estimate for the provincial total and the contribution from each uncertainty source using the multi-stage IPW estimator with and without measurement error are shown in Figure \ref{fig:varbysource}. With both approaches, the biggest contributor to uncertainty is sampling error due to the facilities, followed by sampling error due to the days. Addressing these sources of uncertainty in future survey designs could lead to more precise estimates. A key difference between the approaches with and without measurement error is that the estimated variance contribution from each stage increases when measurement error is considered. We expect this is due to the same reason that the point estimate increases when measurement error is considered - the underlying distribution for each measurement is right-skewed, and simply multiplying the measured emission rates by the mean of the distribution does not adequately account for this. On the provincial level, detection uncertainty is the smallest contributor to uncertainty when measurement error is considered.

\begin{table}[ht]
\centering
\caption{Estimates of the total methane emissions for 2021, variance estimates by uncertainty source, and total variance estimates by stratum using the multi-stage approach with IPW considering all three stages and measurement uncertainty. Due to rounding, small non-zero values are shown as 0.00.  Acronyms: GP = gas plant, MWB = multi-well battery, LNGP = liquid natural gas plant, CO = crude oil, CTF = custom treating facility, GGS = gas gathering system, SWB = single-well battery, MS = meter station, TF = tank farm, NGL = natural gas liquids.}
\label{tab:stratumsummaries}
\begin{tabular}{c|r|rrrr|r}
\multirow{2}{*}{Stratum} & {Total Estimate} &{$\Varone$} & $\Vartwo$ & $\Varthree$ & {$\Varmsmt$} & $\Vptqm$\\ 
&\multicolumn{1}{c|}{(kt/y)} &\multicolumn{4}{c|}{($[\text{kt/y}]^2$)} & ($[\text{kt/y}]^2$) \\ \hline 
  TF and NGL & 0.03 & 0.00 & 0.00 & 0.00 & 0.00 & 0.00 \\ 
  MS & 0.20 & 0.00 & 0.00 & 0.00 & 0.00 & 0.01 \\ 
  Water Hub Battery & 0.22 & 0.00 & 0.00 & 0.05 & 0.04 & 0.09 \\ 
  Gas SWB & 0.23 & 0.01 & 0.01 & 0.00 & 0.01 & 0.03 \\ 
  CO SWB & 0.28 & 0.00 & 0.01 & 0.03 & 0.02 & 0.06 \\ 
  Disposal/storage & 0.39 & 0.04 & 0.02 & 0.00 & 0.01 & 0.07 \\ 
  Mixed OG Battery & 0.40 & 0.00 & 0.01 & 0.00 & 0.01 & 0.02 \\ 
  CTF and GGS & 0.81 & 0.11 & 0.03 & 0.00 & 0.02 & 0.17 \\ 
  CO MWB & 0.82 & 0.00 & 0.10 & 0.01 & 0.02 & 0.13 \\ 
  GP Acid Gas Low & 3.97 & 0.28 & 0.18 & 0.02 & 0.08 & 0.55 \\ 
  Gas MWB (group) & 4.82 & 1.04 & 0.83 & 0.01 & 0.47 & 2.34 \\ 
  Other GP and LNGP & 8.75 & 0.45 & 2.10 & 0.04 & 0.57 & 3.16 \\ 
  Gas MWB (effluent) & 14.71 & 1.41 & 1.61 & 0.09 & 0.53 & 3.64 \\ 
  Wells & 14.75 & 12.19 & 0.35 & 1.42 & 3.24 & 17.20 \\ 
  GP Sweet & 17.14 & 3.97 & 3.47 & 0.05 & 2.42 & 9.92 \\ 
  Compressor Stations & 45.49 & 80.52 & 22.52 & 0.07 & 9.32 & 112.43 \\ \hline
  Population & 113.01 & 100.02 & 31.24 & 1.80 & 17.01 & 150.08 \\ 
   \hline
\end{tabular}
\end{table}

\begin{figure}
    \centering
    \includegraphics[width = 0.9\linewidth]{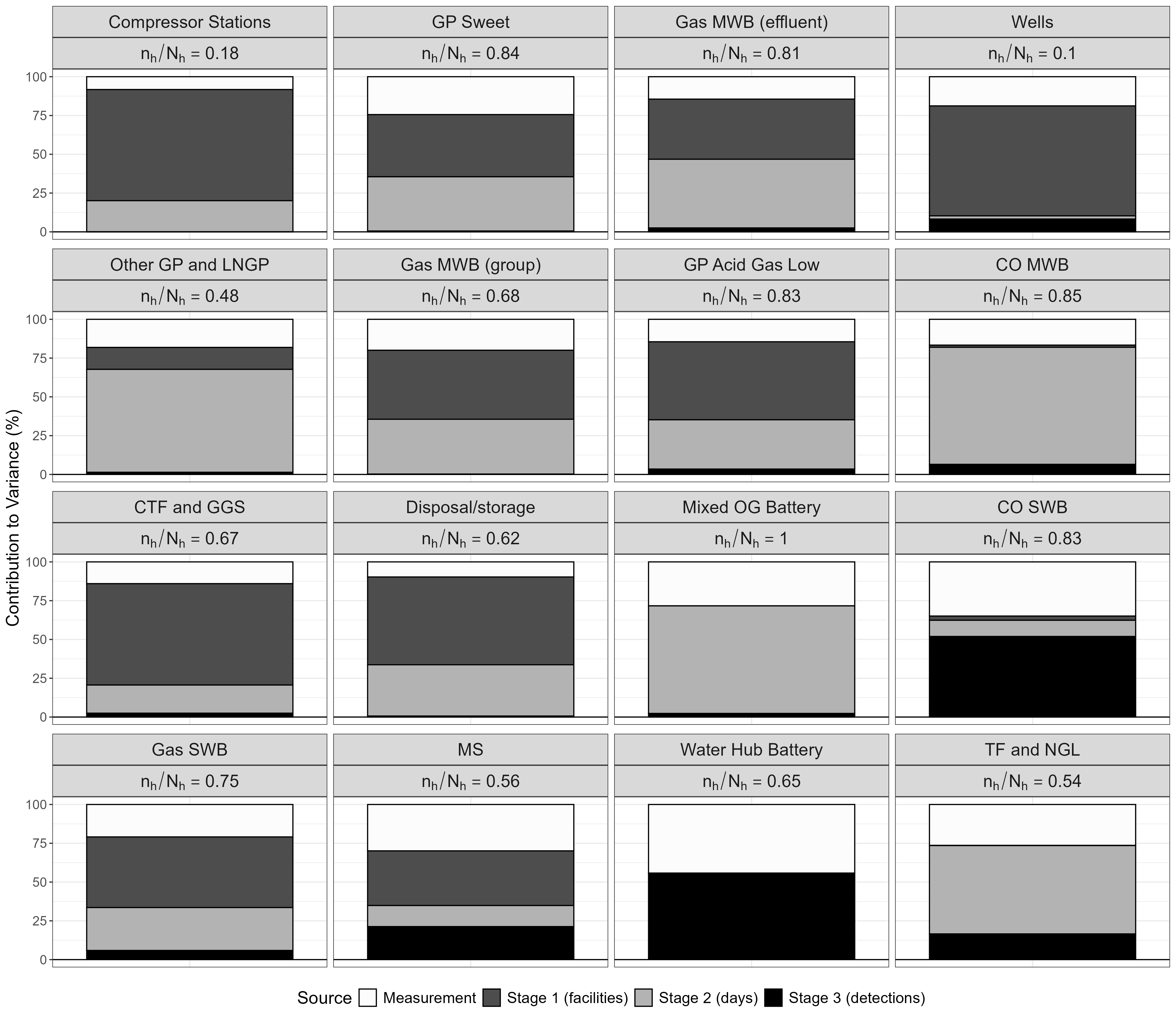}
    \caption{Percent contribution to the estimated total variance by uncertainty source for every stratum in the BC data. Strata are organized from highest (top left) to lowest (bottom right) emitting. The results are from the multi-stage IPW approach which includes stage II and measurement uncertainty. Acronyms: GP = gas plant, MWB = multi-well battery, LNGP = liquid natural gas plant, CO = crude oil, CTF = custom treating facility, GGS = gas gathering system, SWB = single-well battery, MS = meter station, TF = tank farm, NGL = natural gas liquids.}
    \label{fig:varstrata}
\end{figure}

We can investigate the variance breakdowns within individual strata. Figure \ref{fig:varstrata} shows the breakdown of variance contributions from different sources for every stratum. Uncertainty in higher emitting strata tends to be dominated by stage I or stage II. Lower emitting strata are more likely to have measurement uncertainty and detection uncertainty contribute an important amount to the variance. This suggests that reducing the variance in a particular stratum total estimate will require a tailored, case-by-case approach.

\section{Discussion and Conclusions} \label{sec:disc}

We showed how a multi-stage framework can be leveraged to quantify measurable methane emissions given data from a complex aerial survey. We also proposed a MC algorithm which leverages knowledge about the survey technology to incorporate measurement uncertainty into the multi-stage approach. Overall, the methods account for the four key sources of uncertainty in aerial surveys of methane emissions from O\&G: sampling components, sampling days, detection uncertainty, and measurement uncertainty. The results yield similar insights to the nested MC approach that is the current standard for developing measurement-based inventories, with additional abilities to explicitly decompose the variance contribution from different uncertainty sources and a reduced computational burden.

It is important to note that the approach proposed herein only estimates measurable emissions; that is, emissions which have a non-zero detection probability. Estimating emissions which are too small to be detected using GML require alternative data collection and estimation techniques. Methods for estimating unmeasurable emissions are covered in \citet{johnson_creating_2023}. Due to the good performance of the GML technology, measurable emissions have been found to account for a majority of total methane emissions from upstream O\&G in multiple studies of different oil-producing regions \citep{johnson_creating_2023,conrad_measurement-based_2023, conrad_futility_2023}. The presented methods are therefore an important contribution to advancing the ease and accuracy of estimating O\&G methane emissions as a whole. The methods can also be adapted to any methane detection and quantification technology, provided that the probability of detection function and measurement error distribution are available.

A distinction between the methods presented here and the nested MC approach is the way CIs are constructed. The nested MC approach uses quantiles of a bootstrap sample to create CIs whereas here we propose the use of Wald intervals. Bootstrap intervals can be asymmetric reflecting skewed distributions, whereas Wald intervals are symmetric and rely on asymptotic normality for their validity. A strength of the Wald CI is that it is considerably simpler to understand and requires almost no computational effort. Small sample sizes and skewed populations can lead to violations of the normality assumption. We increased the sample size in some strata by merging some facility types, although sample sizes at stages II and III remain small. The sampling distributions shown in \citet{johnson_creating_2023} appear roughly symmetric, reflected in the 95\% CI for the provincial inventory of $-18.2\%$ to $+21.0\%$ (a perfectly symmetric CI would be plus or minus $21\%$, for example). However, it is known that the distribution of emission rates from emitting O\&G facilities is right-skewed.  In a simulation study in SI 1, we found that for one stratum, sweet gas plants, the coverage probability of Wald CIs was noticeably lower than the nominal level. This stratum had an emission rate distribution that departed from normality more than the other strata considered in the simulation, although all followed non-normal distributions. The coverage probability for this stratum was 88\% compared to the nominal 95\%. Still, for most strata, and for the provincial total, the coverage probability was at or close to the nominal level ($> 92\%$) despite the small sample sizes and non-normal emissions distributions, as long as stage II uncertainty was accurately considered. This suggests that the Wald CI has adequate performance for estimating provincial methane emissions from O\&G facilities.

An advantage of the multi-stage approach is its relatively low computational burden compared to the currently used approach. Our approach has reduced computational needs in two ways. First, the currently used approach relies on a complicated MC approach to account for measurement and detection error where numerical integration and root-finding methods are needed at every iteration \citep{johnson_creating_2023}. By addressing detection error using IPW or \Hajek\ estimation, we can significantly simplify the MC algorithm needed to quantify measurement uncertainty. Second, by deriving analytical formulas for estimators and CIs, we remove the need for a bootstrap or other MC approach to be nested within the measurement error MC. The existing nested MC approach requires a total of $B^{MC} \times B^{BS}$ MC iterations, where $B^{MC}$ is the number of iterations in their initial MC and $B^{BS}$ is the number of bootstrap samples taken in the mirror-match bootstrap. Alternatively, since our approach does not rely on a bootstrap procedure, our approach requires only $B$ MC iterations in total, where $B$ is the number of iterations in the measurement uncertainty algorithm. \citet{johnson_creating_2023} require a total of $10^8$ MC iterations to produce their inventory for BC, while we used only $8000$. Additionally, we provide an R package, methaneInventory, which makes the methods described herein accessible, available at \url{https://github.com/augustinewigle/methaneInventory}.

It is unclear whether the use of the \Hajek\ estimator provides benefits over IPW in the context of estimating methane emissions from O\&G. Although it is known that the \Hajek\ estimator is more efficient than IPW in many contexts, the variance estimates for the \Hajek\ estimator were larger than those for IPW in the data analysis for the province and for 10 out of 16 strata (when stage II and measurement uncertainty were considered). It also gives larger point estimates for the totals. Another downside to the \Hajek\ estimator is that it requires manipulation of the stage II and III designs in order to analyse the data, which interferes with the interpretation of the variance decompositions for those stages. Overall, more investigation is needed to understand which stratum and population characteristics lead to efficiency gains from using the \Hajek\ estimator. The IPW approach provides results which are comparable to that of the existing nested MC approach at a much lower computational cost, and results in interpretable variance decompositions, making it an attractive option. Our proposed \Hajek\ approach may be useful in other survey contexts where detection error or non-response leads to empty samples.

In the data analysis, we assume the design at stage II is a SRS of $\ddp$ out of $\D$ days. In reality, the sampling process was more complex. A second day of sampling for a component $p$ was only carried out if there were any detections at the entire site containing $p$ on the initial survey day. This means that the number of days $p$ is sampled, $\ddp$, is actually random, which is not considered in the current variance estimate. The probability that two surveys are conducted at component $p$ is related to the probability of having any detections at the site containing $p$ on day $t_1$, where $t_1$ is the day where the initial survey was conducted for component $p$. This probability is given by
\begin{equation*}
    \gamma_p = 1- \prod_{u \in \text{site} \supset p} \prod_{q = 1}^{Q_{pt_1}}(1- \phi_{pt_1q}).
\end{equation*}
If $\gamma_p$ is approximately zero or one, then $\ddp$ is effectively fixed. We calculated conservative estimates of $\gamma_p$, where the facility was used in place of the site. The first quartile of the estimated $\gamma_p$'s was 1.00, showing that the assumption that $\ddp$ is fixed is reasonable.

Another way in which the data departs from a SRS of days is that sampling was only possible over a relatively short time period. This limits our ability to capture any potential long-term temporal patterns, such as seasonal variation in emission rates. In the SRS, we assume all 365 days have an equal chance of being selected. In reality, only days that are relatively close together have a non-zero probability of being selected, which requires complex modelling to describe. We chose to use a SRS because it is a simple way to incorporate the fact that we want to estimate a quantity that relates to 365 days using only one or two days of data. \citet{johnson_creating_2023} performed a sensitivity analysis where the day-to-day variability was exaggerated via a bootstrap procedure to investigate its impact on the estimate and its uncertainty.  In our approach, more complex and realistic sampling designs could be implemented to provide more accurate uncertainty estimates. Alternatively, in future surveys, sampling could be carried out in a way which more closely resembles SRS and covers a wider temporal range. However, feasibility considerations may prevent the days from being selected using a true probability design. The temporal variations that can be expected at O\&G facilities are still an open question in the literature. The multi-stage approach can be adjusted to incorporate new information as it becomes available.

Similarly to stage II, the sampling at stage I is also impacted by practical constraints which make following any probability design difficult. For example, geographic and accessibility constraints will likely prevent some facilities from being selected, having a true inclusion probability of zero. If the part of the population which has zero probability of being sampled differs from the population with non-zero probability of being sampled, the estimates could be biased. For example, if the northern facilities which could not be sampled due to their remoteness tend to emit more than southern facilities, the results of the survey will underestimate total emissions since the surveyed facilities do not properly represent the population. Analysing the data as if it were a probability sample is in line with the established method where a stratified sampling scheme is assumed in the mirror-match bootstrap \citep{johnson_creating_2023}. It is important to be aware of the potential implications of the simplifying assumptions made when translating the true data collection process to a statistical probability design to be used in the analysis. In future surveys, efforts should be made to clearly define and closely follow sampling procedures to facilitate the analysis. Further, a formal sampling design which accounts for practical considerations could be developed in advance. This would allow for the most accurate analysis.

A key strength of our approach is that it can be easily leveraged in the planning of future surveys. Uncertainty due to the sampling of facilities and days contribute significantly to the uncertainty of population estimates, which are the two sources of uncertainty over which we have the most control, assuming we are not in a position to improve the technical abilities of the technology used to make measurements. The variance contribution from stages I and II can easily be predicted by plugging different sample structures into the true variance formulas for $\Th{}$, given in SI 1, where different representative values for $\Yp$ based on previous data can be used. This gives a simple approach to identify more efficient sampling designs in terms of sampling error, something that is not possible in the existing approach \citep{johnson_creating_2023}.

\clearpage

\appendix

\section{Note on Sample and Population Sizes} \label{sec:samp-pop-sizes}
The information on number of facilities and facility type at each site comes from confidential, anonymized data provided by the MERC and BC OGRIS. This data includes all facilities and wells that had `Active' status in Petrinex, including shut-in, in the province of BC as of June 2021. We use the June 2021 sample sizes in this paper because the information on the number of active facilities of each type at each site was only available to us for June 2021. The survey was conducted in September and October of 2021 and so the sample and population sizes for each facility type reported in the supplementary information of \citet{johnson_creating_2023} were obtained from Petrinex in September and October 2021 respectively. The number of active facilities and wells in the Petrinex database decreased from June 2021 to September 2021, so the sample sizes used to analyse the data in this paper are equal to or greater than those reported in \citet{johnson_creating_2023}. It is worth noting that the decrease in the database does not necessarily mean a decrease in the number of active facilities in the field. For facilities besides wells, the population size was determined using the number of each facility type with Active status in Petrinex in September 2021, when most of the surveys were conducted. For wells, the population and sample sizes were determined using the number of wells with active status in Petrinex as of June 28, 2021 because this represents the breakdown of wells at each site available to us.

\section{Treatment of Wells} \label{sec:wells} 

\citet{johnson_creating_2023} differentiate between on-site wells (wells which share a location with other facilities) and off-site wells (wells at completely separate locations from other facility types). They attribute emissions from on-site wells to other facilities at the site as it may not be possible to distinguish between components associated with the well versus the other facilities, so only off-site wells have emissions attributed to them. We take a different approach because the data from the MERC and BC OGRIS attributed emissions to both on- and off-site wells and does not differentiate between on- and off-site wells.

Unlike other facility types, multiple wells may share the same pieces of equipment. For example, four wells may be connected to a single tank, and thus, if the tank is emitting methane, it cannot be attributed to a single well. For every leaking component that is associated with a well, we sum the estimated average emission rate for that survey occasion for all detected well components at that site and divide by the number of wells at the site, and set 
\begin{equation}
    \Yhpt{} = \frac{\sum_{p \in \SI(\text{Wells})}\Yhpt{}}{\text{Number of wells at site}}
\end{equation}
for all the wells at that site for all $t$. The variance is given by $\sum_{i \in \SI(\text{Wells})}\Vhpt{}/(\text{Number of wells at site})^2$. This is similar to the approach in \citet{johnson_creating_2023}, where emissions are divided equally among wells at the same site.

\section{Competing interests}
No competing interest is declared.

\section{Author contributions statement}

AW and AB conceptualized and developed the methodology. AW wrote the software, cleaned and analysed the data, and drafted the manuscript. AB edited the manuscript and supervised the project. 

\section{Data Availability Statement}

A subset of the data used to produce the methane inventory for BC in 2021 is available in the R package methaneInventory (github.com/augustinewigle/methaneInventory). See SI 2 for details. Please contact Peter Kos (Peter.Kos@gov.bc.ca) from the MERC to request access to the full dataset.

\section{Acknowledgments}
We acknowledge BC OGRIS and the MERC for providing survey data and feedback on an early draft of the manuscript. We gratefully acknowledge Peter Kos for his assistance in interpreting the data and Prof. Kyle J. Daun for bringing statistical issues in methane emissions quantification to our attention. This work was supported by the Natural Sciences and Engineering Research Council of Canada (AW: CGS D-569445-2022, AB: RGPIN-2019-04404).

\bibliographystyle{abbrvnat}
\bibliography{refs}

\end{document}


\maketitle

\tableofcontents

\newcommand{\POD}{\text{POD}}
\newcommand{\Hajek}{H\'{a}jek}
\newcommand{\Unif}{\text{Unif}}

\newcommand{\E}{\text{E}}
\newcommand{\Var}{\text{Var}}
\newcommand{\Cov}{\text{Cov}}

\newcommand{\Ed}{\E_{ptq}}
\newcommand{\Vard}{\Var_{ptq}}

\newcommand{\Eme}{\E_{m}}
\newcommand{\Varme}{\Var_{m}}

\newcommand{\Ymeas}{\dot{Y}}

\newcommand{\stageone}{\text{I}}
\newcommand{\stagetwo}{\text{II}}
\newcommand{\stagethree}{\text{III}}

\newcommand{\UI}{U^{\stageone}}
\newcommand{\UII}{U^{\stagetwo}_p}
\newcommand{\UIII}{U^{\stagethree}_{pt}}

\newcommand{\SI}{S^{\stageone}}
\newcommand{\SII}{S^{\stagetwo}_p}
\newcommand{\SIII}{S^{\stagethree}_{pt}}

\newcommand{\sI}{s^{\stageone}}
\newcommand{\sII}{s^{\stagetwo}_p}
\newcommand{\sIII}{s^{\stagethree}_{pt}}
\newcommand{\sIIstar}{s^{\stagetwo *}_p}

\newcommand{\SIIstar}{S^{\stagetwo *}_p}
\newcommand{\SIIIstar}{S^{\stagethree *}_{pt}}

\newcommand{\D}{D}
\newcommand{\ddp}{d_p}
\newcommand{\Qpt}{Q_{pt}}

\newcommand{\Nhp}{N_{h(p)}}
\newcommand{\nhp}{n_{h(p)}}
\newcommand{\Nhm}{N_{h(l)}}
\newcommand{\nhm}{n_{h(l)}}
\newcommand{\Nh}{N_{h}}
\newcommand{\nh}{n_{h}}

\newcommand{\pip}{\pi_p^\stageone} 
\newcommand{\pim}{\pi_l^\stageone} 
\newcommand{\pipm}{\pi_{pl}^\stageone} 

\newcommand{\pit}{\pi_{t\mid p}^\stagetwo} 
\newcommand{\piu}{\pi_{u\mid p}^\stagetwo} 
\newcommand{\pitu}{\pi_{tu\mid p}^\stagetwo} 

\newcommand{\piq}{\pi_{q \mid pt}^\stagethree} 
\newcommand{\pir}{\pi_{r \mid pt}^\stagethree}
\newcommand{\piqr}{\pi_{qr \mid pt}^\stagethree}

\newcommand{\podq}{\phi_{ptq}}
\newcommand{\podr}{\phi_{ptr}}

\newcommand{\phipt}{\phi_{pt\cdot}} 
\newcommand{\phipu}{\phi_{pu\cdot}} 

\newcommand{\phihpt}{\hat{\phi}_{pt\cdot}} 
\newcommand{\phihir}{\hat{\phi}_{pu\cdot}} 
\newcommand{\podmean}{\mu_{pt\cdot}} 

\newcommand{\sumppop}{\sum_{p \in \UI}} 
\newcommand{\summpop}{\sum_{l \in \UI}} 
\newcommand{\sumtpop}{\sum_{t \in \UII}} 
\newcommand{\sumupop}{\sum_{u \in \UII}} 
\newcommand{\sumqpop}{\sum_{q \in \UIII}} 
\newcommand{\sumrpop}{\sum_{r \in \UIII}}

\newcommand{\sumpsamp}{\sum_{p \in \SI}}
\newcommand{\summsamp}{\sum_{l \in \SI}}
\newcommand{\sumtsamp}{\sum_{t \in \SII}}
\newcommand{\sumusamp}{\sum_{u \in \SII}}
\newcommand{\sumqsamp}{\sum_{q \in \SIII}}
\newcommand{\sumrsamp}{\sum_{r \in \SIII}}

\newcommand{\sumuneq}{\sum_{u \neq t}}

\newcommand{\sumtstarsamp}{\sum_{t \in \SIIstar}}
\newcommand{\sumustarsamp}{\sum_{u \in \SIIstar}}

\newcommand{\Yptq}{Y_{ptq}} 
\newcommand{\Yb}{\overline{Y}}
\newcommand{\Ypt}{\Yb_{pt\cdot}}
\newcommand{\Ypu}{\Yb_{pu\cdot}}
\newcommand{\Yp}{\Yb_{p\cdot\cdot}}
\newcommand{\Ym}{\Yb_{m\cdot\cdot}}

\newcommand{\Yptqmeas}{\Ymeas_{ptq}}

\newcommand{\Yhptq}[1]{\hat{Y}^{\text{#1}}_{ptq}}
\newcommand{\Yhpt}[1]{\hat{\Yb}^{\text{#1}}_{pt\cdot}}
\newcommand{\Yhp}[1]{\hat{\Yb}^{\text{#1}}_{p\cdot\cdot}}

\newcommand{\Yhptl}[1]{\hat{Y}^{\text{#1}}_{ptr}}
\newcommand{\Yhpu}[1]{\hat{\Yb}^{\text{#1}}_{pu\cdot}}
\newcommand{\Yhm}[1]{\hat{\Yb}^{\text{#1}}_{l\cdot\cdot}}

\newcommand{\Th}[1]{\hat{T}^{\text{#1}}{}}

\newcommand{\uptq}{u_{ptq}}
\newcommand{\hptq}{a_{ptq}}
\newcommand{\V}{\text{V}}
\newcommand{\Vh}{\hat{\V}}
\newcommand{\Vtilde}{\tilde{\V}}

\newcommand{\Vtot}[1]{{\V}^{\text{3 stage, #1}}}
\newcommand{\Vone}[1]{{\V}^{\stageone, #1}}
\newcommand{\Vtwo}[1]{{\V}^{\stagetwo, #1}}
\newcommand{\Vthree}[1]{{\V}^{\stagethree, #1}}

\newcommand{\Vhtott}{\hat{\V}^{\text{3 stage}}}
\newcommand{\Vhonee}{\hat{\V}^{\stageone}}
\newcommand{\Vhtwoo}{\hat{\V}^{\stagetwo}}
\newcommand{\Vhthreee}{\hat{\V}^{\stagethree}}

\newcommand{\Vp}{\V_{p\cdot\cdot}}
\newcommand{\Vhp}[1]{\hat{\V}^{\text{#1}}_{p\cdot\cdot}}

\newcommand{\Vpt}{\V_{pt\cdot}}
\newcommand{\Vhpt}[1]{\hat{\V}^{\text{#1}}_{pt\cdot}}

\newcommand{\Varone}{\Vtilde^{\stageone}}
\newcommand{\Vartwo}{\Vtilde^{\stagetwo}}
\newcommand{\Varthree}{\Vtilde^{\stagethree}}
\newcommand{\Varmsmt}{\Vtilde^{m}}
\newcommand{\Vptqm}{\Vtilde^{\text{Total}}}

\newcommand{\tauh}{\tilde{\tau}}

\newcommand{\Ymeasvec}{\bm{\Ymeas}}
\newcommand{\Ytruevec}{\bm{Y}}

\newcommand{\sumb}{\sum_{b = 1}^B}

\newcommand{\sumh}{\sum_{h = 1}^H}
\newcommand{\sumpsampstrata}{\sum_{\substack{p \in \SI, \\ p \in \Omega_{h}}}}
\newcommand{\summsampstrata}{\sum_{\substack{l \in \SI, \\ l \in \Omega_{h}}}}

\newcommand{\sumc}{\sum_{c = 1}^{5000}}

\newcommand{\prodp}{\prod_{p \in \SI}}
\newcommand{\prodtstar}{\prod_{t\in \SIIstar}}
\newcommand{\prodt}{\prod_{t \in \SII}}
\newcommand{\prodtno}{\prod_{t \in \SI\setminus \SIIstar}}

\clearpage
\section{Summary of Notation Used in the Article}

{
\begin{table}[H]
    \centering{\small
        \caption{A summary of the notation and terms used in the article.}
    \label{tab:notation}
    \begin{tabular}{c|p{0.85\linewidth}}
        Notation & Interpretation \\ \hline
        $ptq$ & Represents the three-stage sampling design \\
        $m$ & Represents the measurement model \\
        $\stageone, \stagetwo,\stagethree$ & Represents stage one, stage two, and stage three of the sampling design, respectively \\
        $\hat{}$ & Represents an estimator with respect to the $ptq$-design \\
        $\tilde{}$ & Represents an estimator with respect to the measurement model \\
        $(h)$ & Indicates that the quantity is a property of stratum $h$ \\
        $T; \Th{}$ & The target of estimation from a design-based perspective; its $ptq$-estimator, respectively \\
        $\Yptq; \uptq; \hptq$ & The true emission rate in kg/h, wind speed in m/s, and altitude of the plane in m from component $p$ on day $t$ in pass $q$ \\
        $\Yptqmeas$ & The measured emission rate from component $p$ on day $t$ in pass $q$ \\
        $\Ypt$ & The true average emission rate from component $p$ on day $t$, averaged over the passes \\
        $\Yhpt{}; \Yhpt{IPW}; \Yhpt{\Hajek}$ & A generic $ptq$-estimator of $\Ypt$, the IPW estimator of $\Ypt$, and the \Hajek\ estimator of $\Ypt$ \\
        $\Yp$; $\Yhp{}$ & The true average emission rate from component $p$, averaged over all days and passes; its generic $ptq$-estimator \\
        $\Vtot{}; \Vhtott$& The true variance of $\Th{}$ with respect to the $ptq$-design; its estimator \\
        $\V^{j}; \Vh^{j}$ & The true variance contribution from stage $j$ for $j = \stageone, \stagetwo, \stagethree$; their $ptq$-estimators  \\
        $\Vp; \Vhp{}$ & The true $ptq$-variance of $\Yhp{}$; its $ptq$-estimator \\
        $\Vpt; \Vhpt{}$ & The true $ptq$-variance of $\Yhpt{}$; its $ptq$-estimator \\
        $\UI; \UII; \UIII$ & The population of PSUs (components); The population of SSUs (days) for component $p$; The population of TSUs (passes) for component $p$ and day $t$ \\
        $\SI; \SII; \SIII$ & The stage I sample of components; The stage II sample of days for component $p$; the stage III sample of passes for component $p$ on day $t$ \\
        $\SIIstar$ & The stage II sample of days for which methane was detected at least once; replaces $\SII$ when $\Yhpt{\Hajek}$ is used\\
        $\pip; \pipm$ & The probability that $p$ is selected into $\SI$; the probability that components $p$ and $l$ were selected into $\SI$ \\
        $\pit; \pitu$ & The probability that $t$ is selected into $\SII$ given that $p$ is in $\SI$; the probability that $t$ and $u$ are selected into $\SII$ given that $p$ is in $\SI$ \\
        $\piq; \piqr$ & The probability that pass $q$ is included in $\SIII$, that is, the probability that methane is detected in pass $q$, given that $p$ is in $\SI$ and $t$ is in $\SII$; the probability that methane is detected in passes $q$ and $r$ given  that $p$ is in $\SI$ and $t$ is in $\SII$\\
        ${\pit}^*; {\pitu}^*$ & The probability that $t$ is selected into $\SII$ given that $p$ is in $\SI$ AND methane is detected in at least one pass on day $t$; the probability that $t$ and $u$ are selected into $\SII$ given that $p$ is in $\SI$ AND methane is detected at least once on days $t$ and $u$ \\
        ${\piq}^*; {\piqr}^*$ & The probability that methane was detected in pass $q$ given that $t$ is in $\SIIstar$ and $p$ is in $\SI$; the probability that methane was detected in passes $q$ and $r$ given that $t$ is in $\SIIstar$ and $p$ is in $\SI$ \\
        $\podq$ & The probability of detecting methane from component $p$ on day $t$ in pass $q$. Calculated using Equation (1) \\
        $\phipt; \phihpt$ & The probability of detecting methane at least once from component $p$ on day $t$; its estimate using MI \\
        $\Nhp, \nhp$ & The number of facilities in $\UI$ belonging to the stratum containing component $p$; the number of facilities in $\SI$ belonging to the stratum containing component $p$ \\
        $\D$; $\ddp$ & The number of days in the population; the number of days on which component $p$ is sampled \\
        $\Qpt$ & The number of passes made over component $p$ on day $t$ \\
        $\tau; \tauh$ & The target of estimation considering the $ptq$-design and the measurement model; its estimator \\
        $\V^{\text{Total}}; \Vptqm$ & The true variance of $\tauh$ with respect to $ptq$ and $m$; its estimator \\
        $\V^j; \Vtilde^j$ & The true variance contribution from $j$ with respect to $ptq$ and $m$; its estimator, for $j =m, \stageone, \stagetwo,\stagethree$ 
    \end{tabular}
    }
\end{table}
}
\section{True Variance Formulas for Multi-Stage Approach}\label{sec:truevars}

The true variance of $\Th{}$ with respect to the $ptq$-design for generic sampling designs at stages I and II is given by
\begin{equation}
    \sumppop \sumppop (\pipm - \pip\pim) \frac{\Yp}{\pip}\frac{\Ym}{\pim} + \sumppop \frac{\Vp}{\pip},
\end{equation}
where $\Vp$ is is the design-based variance of $\Yhp{IPW}$ or $\Yhp{\Hajek}$ depending on the approach used. $\Vp^{\text{IPW}}$ is given by
\begin{equation}
    \frac{1}{\D^2}\left\{\sumtpop\sumupop (\pitu-\pit\piu)\frac{\Ypt}{\pit}\frac{\Ypu}{\piu} + \sumtpop \frac{\Vpt^{IPW}}{pit}\right\}
\end{equation}
where
\begin{equation}
    \Vpt^{\text{IPW}} = \frac{1}{\Qpt^2}\sumqpop \frac{(1-\phipt)}{\phipt}\Yptq^2
\end{equation}
when the non-detection follows Poisson sampling. For the \Hajek\ estimator, $\Vp^{\text{\Hajek}}$ is given by
\begin{equation}
    \frac{1}{\D^2}\left\{\sumtpop\sumupop ({\pitu}{}^*-{\pit}{}^*{\piu}{}^*)\frac{\Ypt}{{\pit}{}^*}\frac{\Ypu}{{\piu}{}^*} + \sumtpop \frac{\Vpt^{\text{\Hajek}}}{{\pit}{}^*}\right\}
\end{equation}
where 
\begin{equation}
    \Vpt^{\text{\Hajek}} \approx \frac{1}{\Qpt^2}\left\{\sumqpop \phipt\frac{(1-\podq)}{\podq}(\Yptq - \Ypt)^2 + (\phipt - 1)\left(\sumqpop (\Yptq-\Ypt)\right)^2\right\}
\end{equation}
is a linear approximation of the variance using a Taylor series expansion.

\clearpage

\section{Proof of Equivalency of Sampling Designs}\label{sec:equiv}

We refer to the original design as the design where the stage II sample is governed by inclusion probabilities $\pit$ and $\pitu$ and the stage III sample is governed by $\piq$ and $\piqr$. On the other hand, the modified design is the design where the stage II and III samples are governed by $\pit{}^*$, $\piqr{}^*$, $\piq{}^*$, and $\piqr{}^*$.

Let $\sI$ be a fixed but arbitrary sample of PSUs.
Similarly, let $\sII$ be a fixed but arbitrary sample of SSUs for all $p \in \sI$, and let $\sIII$ be a fixed but arbitrary sample of TSUs for all $t \in \sII$ and $p \in \sI$.
Together, let $s = \{p \in \sI, t \in \sII, q \in \sIII\}$ represent a fixed but arbitrary sample of PSUs, SSUs and TSUs.
To show equivalence between the original and modified design, we will show that the probability of observing $s$ is equal under both designs.
 
Under the original design, the probability of observing $s$ is
\begin{align*}
    \Pr(\SI = &~\sI, \SII = \sII \text{ for all } p \in \sI, \SIII = \sIII \text{ for all } p \in \sI \text{ and } t \in \sII) \\
&= \Pr(\SI = \sI) \left[ \prod_{p \in \sI} \Pr(\SII = \sII \mid \SI = \sI) \left\{ \prod_{t \in \sII} \Pr(\SIII = \sIII \mid \SI = \sI, \SII = \sII) \right\} \right].
\end{align*}

Under the modified design, let $\sIIstar = \{t \in \sII  \text{ s.t. } Z_{pt} = 1\}$ where $Z_{pt} = 1$ if there are any detections at component $p$ on day $t$ and zero otherwise. Then the probability of $s$ in the modeified design is have
\begin{align*}
    \Pr(\SI = &~\sI, \SIIstar = \sIIstar \text{ for all } p \in \sI, \SIII = \sIII \text{ for all } p \in \sI \text{ and } t \in \sIIstar) \\
    &= \Pr(\SI = \sI) \left[ \prod_{p \in \sI} \Pr(\SIIstar = \sIIstar \mid \SI = \sI) \left\{ \prod_{t \in \sIIstar} \Pr(\SIII = \sIII \mid \SI = \sI, \SIIstar = \sIIstar) \right\} \right] \\
    &= \Pr(\SI = \sI) \left[\prod_{p \in \sI} \Pr(\SII = \sII \mid \SI = \sI) \left\{\prod_{t \in \sIIstar} \phipt\right\} \left\{\prod_{t \in \sII \setminus \sIIstar} (1-\phipt)\right\} \right. \\ 
     & \left.\hspace{7cm}\left\{\prod_{t \in \sIIstar} \Pr(\SIII = \sIII \mid \SI = \sI, \SIIstar = \sIIstar)\right\}\right] \\
    &=  \Pr(\SI = \sI) \left[\prod_{p \in \sI} \Pr(\SII = \sII \mid \SI = \sI) \left\{\prod_{t \in \sIIstar} \phipt\right\} \left\{\prod_{t \in \sII \setminus \sIIstar} (1-\phipt)\right\}\right. \\
    &\hspace{7cm} \left.\left\{\prod_{t \in \sIIstar} \frac{\Pr(\SIII = \sIII \mid \SI = \sI, \SII = \sII)}{\phipt}\right\} \right] \\
    &= \Pr(\SI = \sI) \left[\prod_{p \in \sI} \Pr(\SII = \sII \mid \SI = \sI) \left\{\prod_{t \in \sII \setminus \sIIstar} (1-\phipt)\right\} \left\{\prod_{t \in \sIIstar} \Pr(\SIII = \sIII \mid \SI = \sI, \SII = \sII)\right\}\right] \\
    &= \Pr(\SI = \sI) \left[ \prod_{p \in \sI} \Pr(\SII = \sII \mid \SI = \sI) \left\{ \prod_{t \in \sII} \Pr(\SIII = \sIII \mid \SI = \sI, \SII = \sII) \right\} \right],
\end{align*}
since for $t \in \sIIstar$, 
\begin{align*}
    \Pr(\SIII = \sIII \mid \SI = \sI, \SIIstar = \sIIstar) &= \Pr(\SIII = \sIII \mid \SI = \sI, \SII = \sII, Z_{pt} = 1) \\
    &= \frac{\Pr(\SIII = \sIII \cap Z_{pt} = 1 \mid \SI = \sI, \SII = \sII)}{\Pr(Z_{pt} = 1 \mid \SI = \sI, \SII = \sII)} \\
    &= \frac{\Pr(\SIII = \sIII \mid \SI = \sI, \SII = \sII)}{\phipt},
\end{align*}
and for $t \notin \sIIstar$,
\begin{equation*}
    \Pr(\SIII = \sIII \mid \SI = \sI, \SII = \sII) = (1-\phipt).
\end{equation*}
Thus, the original design is equivalent to the modified design.

\clearpage

\section{Derivation of Equations in Section 3.4.2}

\subsection{Variance Estimator for the IPW Approach with SRS}
 
\begin{align*}
    \Vhp{IPW} &= \frac{1}{\D^2}\left\{\sumtsamp\sumusamp \frac{(\pitu - \pit\piu)}{\pitu}\frac{\Yhpt{}}{\pit}\frac{\Yhpu{}}{\piu} + \sumtsamp \frac{\Vhpt{}}{\pit}\right\}\\
    &= \frac{1}{\D^2}\left\{\sumtsamp \frac{(1-\pit)}{(\pit)^2}(\Yhpt{IPW})^2 + \sumtsamp \sumuneq \frac{\pitu-\pit\piu}{\pitu}\frac{\Yhpt{IPW}}{\pit}\frac{\Yhpu{IPW}}{\piu} + \sumtsamp \frac{\Vhpt{}}{\pit}\right\} \\
    &= \frac{1}{\D^2}\left\{\sumtsamp \left(\frac{\D}{\ddp}\right)^2\left(1-\frac{\ddp}{\D}\right)(\Yhpt{IPW})^2 + 
    \sumtsamp \sumuneq\frac{\frac{\ddp}{\D}\left(\frac{\ddp-1}{\D-1}\right) - \left(\frac{\ddp}{\D}\right)^2}{\frac{\ddp}{\D}\left(\frac{\ddp-1}{\D-1}\right)} \left(\frac{\D}{\ddp}\right)^2\Yhpt{IPW} \Yhpu{IPW}  +\sumtsamp\frac{\D}{\ddp}\Vhpt{IPW}\right\} \\
    &= \frac{1}{\D^2}\left\{\sumtsamp \frac{\D(\D-\ddp)}{\ddp^2}(\Yhpt{IPW})^2 + \sumtsamp\sumuneq \frac{\D(\ddp-\D)}{\ddp^2(\ddp-1)}\Yhpt{IPW}\Yhpu{IPW} + \sumtsamp \frac{\D}{\ddp} \Vhpt{IPW}\right\}\\
    &= \frac{1}{\D^2}\left[\sumtsamp \left\{\frac{\D(\D-\ddp)}{\ddp^2} - \frac{\D(\ddp-\D)}{\ddp^2(\ddp-1)}\right\}(\Yhpt{IPW})^2 +
    \sumtsamp\sumusamp \frac{\D(\ddp-\D)}{\ddp^2(\ddp-1)} \Yhpt{IPW}\Yhpu{IPW} + \sumtsamp \frac{\D}{\ddp} \Vhpt{IPW}\right] \\
    &= \frac{1}{\D^2}\left\{\sumtsamp \frac{\D(\D-\ddp)}{\ddp(\ddp-1)}(\Yhpt{IPW})^2 + \frac{\D(\ddp-\D)}{\ddp^2(\ddp-1)}\sumtsamp\sumusamp \Yhpt{IPW}\Yhpu{IPW} \sumtsamp \frac{\D}{\ddp} \Vhpt{IPW}\right\} \\
    &= \frac{1}{\D^2}\left\{\sumtsamp\frac{(\D-\ddp)\D}{\ddp(\ddp-1)}(\Yhpt{IPW})^2 + \frac{\D(\ddp-\D)}{\ddp^2(\ddp-1)}\left(\sumtsamp\Yhpt{IPW}\right)^2 + \sumtsamp\frac{\D}{\ddp}\Vhpt{IPW}\right\}.
\end{align*}

\subsection{Variance Estimator for the \Hajek\ Approach with SRS}

\begin{align*}
    \Vhp{\Hajek} &= \frac{1}{\D^2}\left\{\sumtstarsamp\sumustarsamp \frac{(\pitu{}^*-\pit{}^*\piu{}^*)}{\pitu{}^*}\frac{\Yhpt{\Hajek}}{\pit{}^*}\frac{\Yhpu{\Hajek}}{\piu{}^*} + \sumtsamp \frac{\Vhpt{\Hajek}}{\pit{}^*}\right\} \\
    &= \frac{1}{\D^2}\left\{\sumtstarsamp \frac{1-\pit{}^*}{(\pit{}^*)^2} (\Yhpt{\Hajek})^2 + \sumtstarsamp\sumuneq \frac{\pitu{}^* - \pit{}^*\piu{}^*}{\pitu{}^*}\frac{\Yhpt{\Hajek}}{\pit{}^*}\frac{\Yhpu{\Hajek}}{\piu{}^*} + \sumtsamp \frac{\Vhpt{\Hajek}}{\pit{}^*}\right\} \\
    &= \frac{1}{\D^2}\left\{\sumtstarsamp \frac{1-\pit\phipt}{(\pit\phipt)^2}(\Yhpt{\Hajek})^2 + \sumtstarsamp\sumuneq \frac{(\pit\phipt\phipu-\pit\phipt\piu\phipu)}{\pitu\phipt\phipu} \frac{\Yhpt{\Hajek}}{\pit\phipt}\frac{\Yhpu{\Hajek}}{\piu\phipu} + \sumtstarsamp \frac{\Vhpt{\Hajek}}{\pit\phipt} \right\} \\
    &= \frac{1}{\D^2}\left\{\sumtstarsamp\frac{\left(1-\frac{\ddp}{\D}\phipt\right)}{\left(\frac{\ddp}{\D}\phipt\right)^2}(\Yhpt{\Hajek})^2 + \sumtstarsamp \sumuneq \frac{\frac{\ddp}{\D}\frac{(\ddp-1)}{(\D-1)} - \left(\frac{\ddp}{\D}\right)^2}{\left(\frac{\ddp}{\D}\right)^3\frac{(\ddp-1)}{(\D-1)}} \frac{\Yhpt{\Hajek}}{\phipt}\frac{\Yhpu{\Hajek}}{\phipu} + \sumtstarsamp \frac{\D}{\ddp \phipt}\right\} \\
    &= \frac{1}{\D^2}\left[\sumtstarsamp \frac{\D\{\D-1-\phihpt(\ddp-1)\}}{\ddp(\ddp-1)}\left(\frac{\Yhpt{\Hajek}}{\phipt}\right)^2 + \frac{\D(\ddp - \D)}{\ddp^2(\ddp-1)}\left(\sumtstarsamp \frac{\Yhpt{\Hajek}}{\phipt}\right)^2 + \sumtstarsamp \frac{\D}{\ddp \phipt} \Vhpt{\Hajek}\right].
\end{align*}

\newpage 

\section{Derivation of Total Variance of $\tauh$} \label{sec:msmt-error-deriv}

We will use the following properties:
\begin{enumerate}
    \item $\E(\Th{}^b) = \tau$ for all $b = 1, \dots, B$,
    \item $\Eme(\Th{}^b) = \Th{}$ for all $b = 1, \dots, B$,
    \item $\Th{}^b$ is independent of $\Th{}^c$ with respect to measurement error for $b \neq c$,
    \item $\Eme(\Th{}^b\Th{}^c) = \Eme((\Th{}^b)^2)$ for $b \neq c$,
\end{enumerate}
where properties 2 and 3 give property 4.

The variance of $\tauh$ with respect to measurement error and the design is given by
\begin{align*}
    \Var(\tauh) &= \Var\left(\frac{\sumb \Th{}^b}{B}\right) \\
    &= \frac{1}{B^2}\left(\sumb \Var(\Th{}^b) + \sumb \sum_{c\neq b} \Cov(\Th{}^b, \Th{}^c)\right) \\
    &= \frac{1}{B^2}\left(\sumb \Var(\Th{}^b) + \sumb \sum_{c\neq b} \E(\Th{}^b\Th{}^c) - \E(\Th{}^b)\E(\Th{}^c)\right) \\
    &= \frac{1}{B^2}\left(\sumb \Var(\Th{}^b) + \sumb \sum_{c\neq b} \Ed\Eme(\Th{}^b\Th{}^c) - \E(\Th{}^b)^2\right) \quad \text{using property 1,} \\
    &= \frac{1}{B^2}\left(\sumb \Var(\Th{}^b) + \sumb \sum_{c\neq b} \E[(\Th{}^b)^2] - \E(\Th{}^b)^2\right) \quad \text{using property 4,}\\
    &= \frac{1}{B^2} \left\{B\Var(\Th{}^b) + B(B-1) \Var(\Th{}^b)\right\} \\
    &= \Var(\Th{}^b) \\
    &= \Varme\Ed(\Th{}^b) + \Eme\Vard(\Th{}^b) \\
    &= \Varme (T^b) + \Eme(\V^b).
\end{align*}
We can therefore estimate $\Var(\tauh)$ with
\begin{equation*}
    \frac{\sumb (\Th{}{}^b - \tauh)^2}{B-1} + \frac{\sumb \Vhtott{}^b}{B} 
\end{equation*}
or using the fact that the estimator $\Vhtott{}$ is the sum of variance contributions from each stage,
\begin{equation*}
    \frac{\sumb (\Th{}{}^b - \tauh)^2}{B-1} + \frac{\sumb \Vhonee{}^b}{B}+ \frac{\sumb \Vhtwoo{}^b}{B} + \frac{\sumb \Vhthreee{}^b}{B}.
\end{equation*}

\clearpage

\section{Simulation Study for Wald Confidence Interval Performance}\label{sec:sim}

The aerial survey data collected in BC has several features which may challenge the normality assumption which is relied upon in the Wald CI formula. The sample sizes at stages II and III are very small and the distribution of emission rates are non-normal. Bootstrap CI approaches can circumvent normality assumptions, however they add complexity and computational burden. In support of our aim to provide accessible methods for estimating methane inventories, we perform a simulation study to investigate the properties of the proposed estimators and CIs under similar conditions to the BC data.

\subsection{Data Generation and Sampling}

We created an artificial population, $\UI$, of facilities. We selected four strata from the BC dataset which represented different emission rate distributions and sampling fractions. These strata were crude oil single-well batteries (83\% of facilities represented in $\SI$, fifth lowest emitting stratum in data analysis), meter stations (56\% of facilities represented in $\SI$, third lowest emitting stratum), sweet gas plants (84\% of facilities represented in $\SI$, second highest emitting stratum), and compressor stations (18\% of facilities represented in $\SI$, highest emitting stratum). We use the same population sizes $\Nh$ as we have in the real population, shown in Table 1. Then, a number of components was randomly assigned to each facility in the artificial population, drawn from a discrete uniform distribution from one to 50. Each component was assigned as either emitting or not, drawn from a Bernoulli distribution with a probability of emitting equal to 0.055. These two distributions were chosen so that the simulated population had a similar true total to the estimated total from the four strata in the data analysis.

Next, for 365 days, emitting components were assigned a daily mean emission rate drawn from a stratum-specific distribution. To obtain stratum-specific emission rate distributions, the non-zero emission rates in the motivating dataset were bias-corrected as suggested in Equation (3). Then, a log-normal distribution was fit to the bias-corrected emission rates in the motivating dataset for each stratum to give unique log-normal distributions. Distributions were fit using the fitdistrplus package in R using the moment-matching method. The daily emission rate for component $p$ was sampled from the log-normal distribution corresponding to the stratum of component $p$. Note that this is not exactly the same quantity as $\Ypt$, where the mean is taken over the discrete number of passes.

Once daily average emission rates were assigned, for each emitting component and day, a number of passes is randomly assigned by drawing from the distribution of passes in the observed data. An instantaneous emission rate $\Yptq$, altitude $\hptq$, and wind speed $\uptq$ are assigned at each pass. The height and wind speed are drawn from normal distributions with mean and standard deviation equal to those observed in the data. The wind speed distribution is truncated at zero. Emission rates are drawn from a normal distribution with mean equal to the previously assigned daily mean and standard deviation proportional to the daily mean. The proportion is determined by taking the median of the ratio between mean and standard deviation observed in the data, by strata.

Sampling was carried out following the three-stage sampling design described in Section 3.4. First, a simple random sample of $\nh$ facilities was selected out of $\Nh$ in the artificial population. All components belonging to  selected facilities are part of $\SI$, the stage I sample. Next, a SRS of two out of 365 days was selected for each component in $\SI$ to make up $\SII$. Finally, for all $p \in \SI$ and $t \in \SII$, detections were simulated in each pass $p = 1, \dots, \Qpt$ according to a Bernoulli distribution where the success probability is calculated using Equation (1) with $\Yptq$, $\uptq$, and $\hptq$. The passes with successful detections form $\SIII$.  

For each sample, the stratum total was estimated using two types of estimators and two assumed stage II designs, giving four different estimation approaches in all. Either IPW or \Hajek\ estimators were used, assuming a stratified SRS at stage I, and for stage II, either a SRS of two out of 365 or a SRS of two out of two (ignoring stage II uncertainty) as described in Section 3.4. All estimation methods follow those described in Sections 3.2 and 3.3. The population total was estimated by summing the strata totals. 95\% CIs for individual strata and for the population were calculating using the Wald method described in Equation (8). The sampling and estimation process was carried out 5000 times, giving a sample of 5000 stratum-level and overall population total estimates, denoted $\Th{}(h)^c$ and $\Th{}{}^c$ for $c = 1, \dots, 5000$. The bias, variance, and mean-squared-error (MSE) of $\Th{}$ for the four different methods were estimated according to
\begin{equation*}
    \text{Bias}(\Th{}) = \left(\frac{\sumc \Th{}{}^c}{5000}\right) - T,
\end{equation*}
where $T$ is the true value calculated for the artificial population,
\begin{equation*}
    \Var(\Th{}) = \frac{\sumc (\Th{}{}^c - \frac{\sumc \Th{}{}^c}{5000})^2}{5000-1}
\end{equation*}
and
\begin{equation}
    \text{MSE}(\Th{}) = \Var(\Th{}) + \left\{\text{Bias}(\Th{})\right\}^2.
\end{equation} 
Percent bias is calculated by dividing the bias by the true value $T$ and multiplying by 100\%. The coverage probability of 95\% CIs were calculated according to
\begin{equation}
    \text{Coverage probability}(\Th{}) = \frac{\sumc \mathcal{I}_{T \in (l^c, u^c)}}{5000}
\end{equation}
where $(l^c, u^c)$ represents the lower and upper bounds of the 95\% Wald CI calculated for the $c^{\text{th}}$ estimate, $\Th{}{}^c$. The bias, MSE and coverage probability for the stratum-level estimator $\Th{}(h)$ were calculated similarly.

\subsection{Results}

\begin{figure}
    \centering
    \includegraphics[width = 0.95\linewidth]{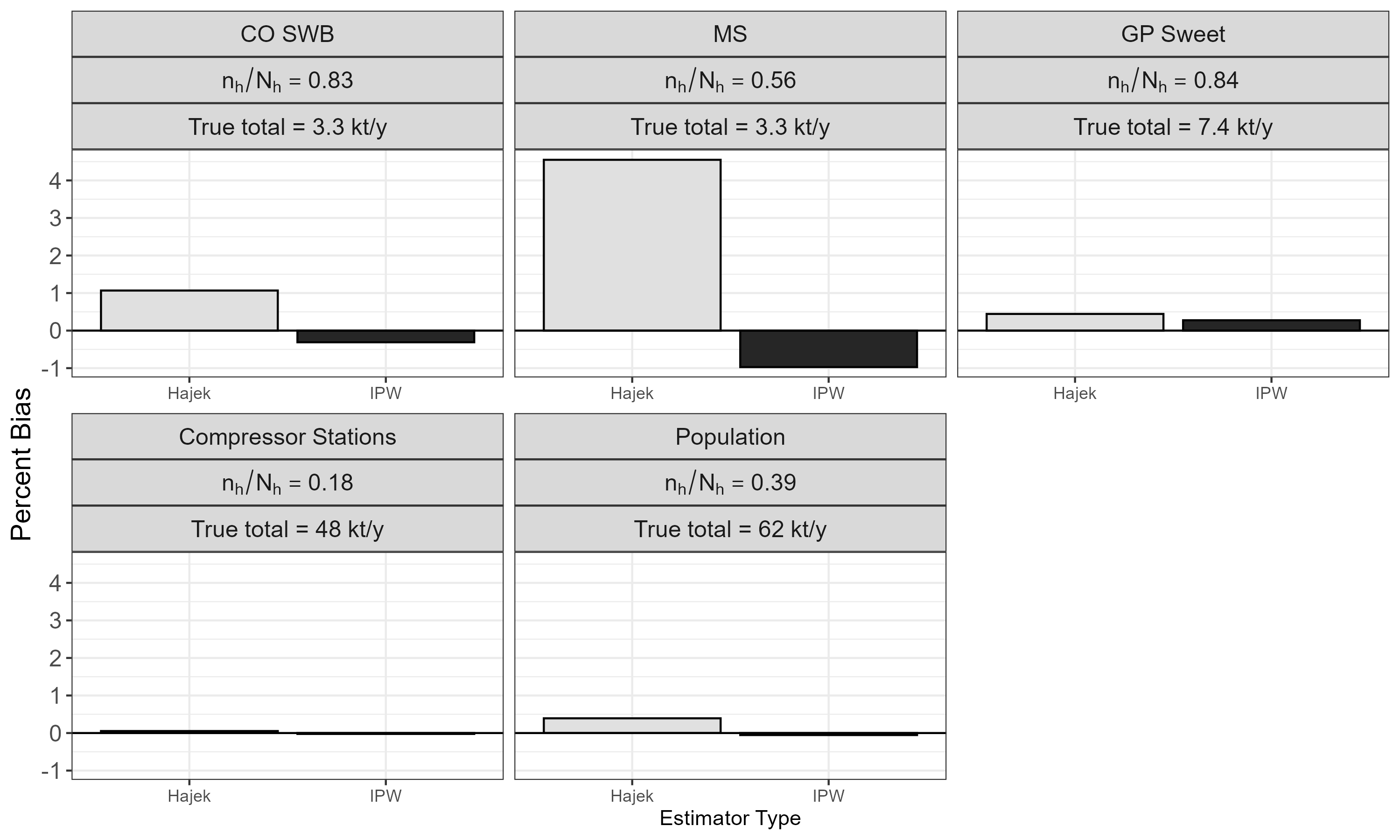}
    \caption{Percent bias for different estimators of the stratum totals $T(h)$ and provincial total $T$.}
    \label{fig:sim-strata-bias}
\end{figure}

The percent bias of the estimators for the four strata and the artificial population are shown in Figure \ref{fig:sim-strata-bias}. We differentiate only between the estimator type (IPW or \Hajek) and not between assumed stage II design in showing these results because the different designs only affect the variance, not the point estimate. For all strata and the population, the \Hajek\ estimator results in a larger magnitude of bias than using IPW, and tends to be biased upwards. The largest magnitude of percent bias is seen when estimating the total for meter stations, at 4.2\% bias. The percent bias tends to be larger for strata with lower true totals. This is likely due to the lower detection probabilities, leading to smaller sample sizes at stage III. For the population total, both approaches have low bias below 0.5\%.

\begin{figure}
    \centering
    \includegraphics[width=0.95\linewidth]{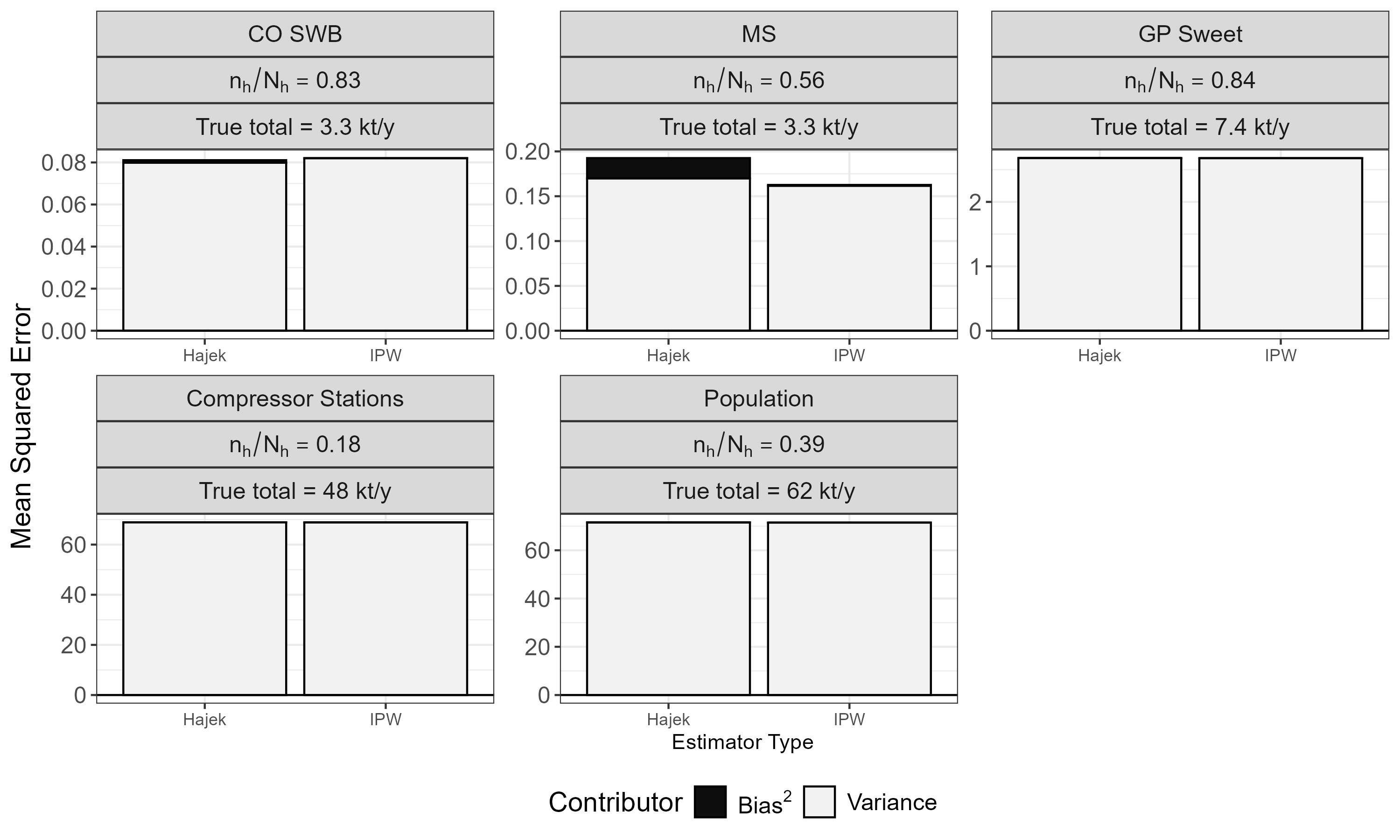}
    \caption{Mean squared error for different estimators of the stratum totals $T(h)$ and provincial total $T$ when considering stage II uncertainty, including the contribution from bias and from variance.}
    \label{fig:sim-strata-mse}
\end{figure}

The MSE results of the simulation study are shown in Figure \ref{fig:sim-strata-mse} for the approaches which considered uncertainty at stage II. The MSE are very similar between the \Hajek\ and IPW approaches for all strata, except for the meter stations stratum, where IPW has noticeably lower MSE. In all cases, the biggest contributor to the MSE is the variance. Despite the \Hajek\ estimator being more efficient than IPW in many situations, in this simulation study we see little to no efficiency gains in using the \Hajek\ estimator. Conclusions are similar for the approaches which did not consider stage II uncertainty. 

\begin{figure}
    \centering
    \includegraphics[width = 0.95\linewidth]{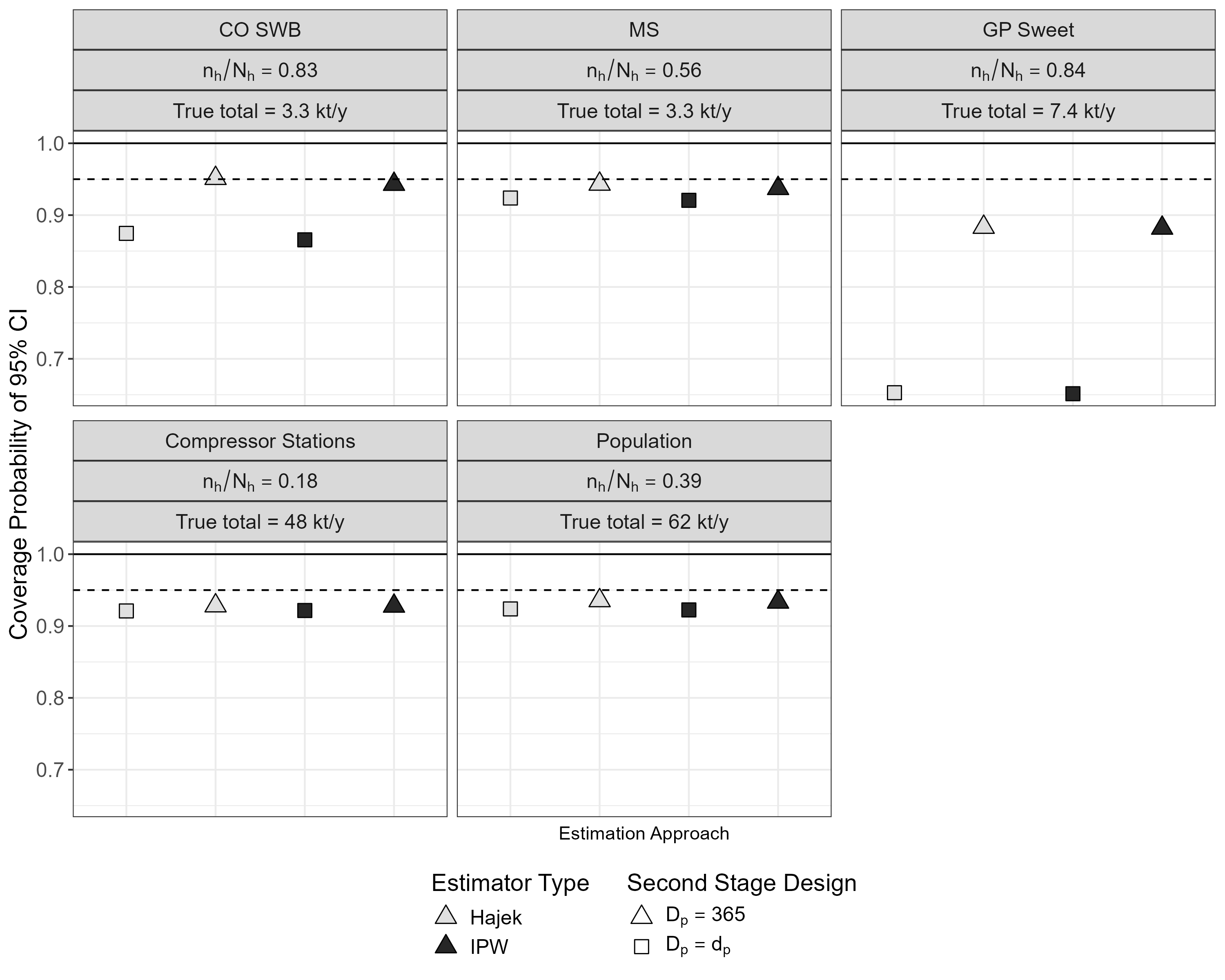}
    \caption{Coverage probability of 95\% Wald CIs for all estimation approaches by stratum. The dotted line represents the nominal coverage probability.}
    \label{fig:sim-strata-cp}
\end{figure}

The coverage probabilities of the estimation approaches are shown in Figure \ref{fig:sim-strata-cp}. The approaches which used the true stage II sampling design ($\D = 365$) give coverage the closest to the nominal level. The differences between IPW and \Hajek\ are marginal. When considering the true stage II design, the coverage probability is at the nominal level for crude-oil single well batteries, only slightly below 95\% for meter stations, compressor stations, and the population total, and below 90\% for sweet gas plants. Ignoring the stage II uncertainty by setting $\D = \ddp$ in the analysis leads to reduced coverage probability. This is especially pronounced for the sweet gas plants stratum, which has the lowest coverage probabilities in general. This is likely due to the underlying emission distribution of this stratum. The log-normal distribution fit to the sweet gas plants data had the largest variance out of four stratum included in the simulation, which leads to skewness in the log-normal distribution, which could explain why the performance is worse for this stratum. 

\subsection{Conclusions}

In the simulation study, we found little difference between the \Hajek\ and IPW approaches. The \Hajek\ produced upwards-biased estimates with similar variance to the IPW approach, although the variance was slightly lower for one very low-emitting stratum. We found that Wald CIs perform adequately for most strata and for the population total. However, coverage is reduced as the underlying distribution of emission rates departs from normality, such as in the case of sweet gas plants, where the coverage dropped to 88\% even when considering stage II uncertainty. This is important to keep in mind as it is known that the distribution of emitters in upstream O\&G is highly skewed \citep{brandt_methane_2016, zavala-araiza_methane_2018}. Despite this, the Wald intervals are robust to some levels of non-normality since the coverage probabilities remain relatively close to 95\% for all other strata and the population when stage II uncertainty was considered. The most important factor influencing the performance of the CIs was whether stage II uncertainty was adequately considered. Ignoring stage II uncertainty leads to under-coverage.

\newpage 

\section{Additional Data Analysis Results}

\subsection{Results for Multi-Stage \Hajek\ with Stage II Uncertainty and Measurement Error}

\begin{table}[H]
\centering
\caption{Estimates of the total methane emissions for 2021, variance estimates by uncertainty source, and total variance estimates by stratum}
\begin{tabular}{c|c|cccc|c}
\multirow{2}{*}{Stratum} & {Total Estimate} &{$\Varone$} & $\Vartwo$ & $\Varthree$ & {$\Varmsmt$} & $\Vtilde$\\ 
&(kt/y) &\multicolumn{4}{c|}{($[\text{kt/y}]^2$)} & ($[\text{kt/y}]^2$) \\ \hline 
Population & 118.981 & 105.511 & 34.328 & 0.040 & 17.502 & 157.380 \\ 
  CO SWB & 0.333 & 0.002 & 0.046 & 0.000 & 0.024 & 0.071 \\ 
  Gas SWB & 0.233 & 0.013 & 0.009 & 0.000 & 0.006 & 0.028 \\ 
  Gas MWB (group) & 5.737 & 1.798 & 0.899 & 0.000 & 0.699 & 3.397 \\ 
  Gas MWB (effluent) & 14.958 & 1.496 & 1.711 & 0.003 & 0.518 & 3.728 \\ 
  Mixed OG Battery & 0.392 & 0.000 & 0.014 & 0.000 & 0.006 & 0.019 \\ 
  Water Hub Battery & 0.216 & 0.028 & 0.001 & 0.000 & 0.040 & 0.069\\ 
  GP Sweet & 19.275 & 5.094 & 3.964 & 0.001 & 2.615 & 11.673 \\ 
  GP Acid Gas Low & 4.331 & 0.307 & 0.206 & 0.000 & 0.086 & 0.599 \\ 
  Compressor Stations & 46.937 & 83.106 & 24.170 & 0.007 & 9.852 & 117.134 \\ 
  CO MWB & 0.942 & 0.004 & 0.112 & 0.000 & 0.024 & 0.139 \\ 
  CTF and GGS & 0.797 & 0.105 & 0.034 & 0.001 & 0.023 & 0.162 \\ 
  Disposal/storage & 0.456 & 0.057 & 0.026 & 0.000 & 0.008 & 0.092 \\ 
  MS & 0.231 & 0.005 & 0.006 & 0.000 & 0.004 & 0.015 \\ 
  Other GP and LNGP & 9.349 & 0.791 & 2.098 & 0.001 & 0.615 & 3.506 \\ 
  TF and NGL & 0.030 & 0.000 & 0.001 & 0.000 & 0.000 & 0.002 \\ 
  Wells & 14.765 & 12.707 & 1.031 & 0.028 & 2.812 & 16.576\\ 
   \hline
\end{tabular}
\end{table}

\subsection{Results for Multi-stage IPW with No Stage II Uncertainty and Measurement Error}

\begin{table}[H]
\centering
\caption{Estimates of the total methane emissions for 2021, variance estimates by uncertainty source, and total variance estimates by stratum}
\begin{tabular}{c|c|cccc|c}
\multirow{2}{*}{Stratum} & {Total Estimate} &{$\Varone$} & $\Vartwo$ & $\Varthree$ & {$\Varmsmt$} & $\Vtilde$\\ 
&(kt/y) &\multicolumn{4}{c|}{($[\text{kt/y}]^2$)} & ($[\text{kt/y}]^2$) \\ \hline 
Population & 113.007 & 120.684 & 0.445 & 1.804 & 17.011 & 139.944 \\ 
  CO SWB & 0.283 & 0.002 & 0.007 & 0.033 & 0.022 & 0.064 \\ 
  Gas SWB & 0.233 & 0.015 & 0.001 & 0.002 & 0.006 & 0.022 \\ 
  Gas MWB (group) & 4.820 & 1.299 & 0.003 & 0.006 & 0.466 & 1.774 \\ 
  Gas MWB (effluent) & 14.709 & 1.698 & 0.010 & 0.092 & 0.527 & 2.327 \\ 
  Mixed OG Battery & 0.395 & 0.000 & 0.000 & 0.000 & 0.006 & 0.006 \\ 
  Water Hub Battery & 0.216 & 0.000 & 0.000 & 0.050 & 0.040 & 0.090 \\ 
  GP Sweet & 17.143 & 4.526 & 0.010 & 0.054 & 2.424 & 7.015 \\ 
  GP Acid Gas Low & 3.975 & 0.304 & 0.002 & 0.019 & 0.080 & 0.405 \\ 
  Compressor Stations & 45.493 & 98.791 & 0.318 & 0.069 & 9.318 & 108.496 \\ 
  CO MWB & 0.821 & 0.016 & 0.001 & 0.009 & 0.022 & 0.048 \\ 
  CTF and GGS & 0.805 & 0.118 & 0.002 & 0.004 & 0.023 & 0.147 \\ 
  Disposal/storage & 0.390 & 0.047 & 0.000 & 0.000 & 0.007 & 0.055 \\ 
  MS & 0.197 & 0.005 & 0.002 & 0.003 & 0.004 & 0.014 \\ 
  Other GP and LNGP & 8.749 & 1.449 & 0.033 & 0.043 & 0.573 & 2.099 \\ 
  TF and NGL & 0.030 & 0.000 & 0.000 & 0.000 & 0.000 & 0.001 \\ 
  Wells & 14.748 & 12.415 & 0.055 & 1.420 & 3.237 & 17.127 \\ 
\end{tabular}
\end{table}

\subsection{Results for Multi-stage \Hajek\ with No Stage II Uncertainty and Measurement Error}

\begin{table}[H]
\centering
\caption{Estimates of the total methane emissions for 2021, variance estimates by uncertainty source, and total variance estimates by stratum}
\begin{tabular}{c|c|cccc|c}
\multirow{2}{*}{Stratum} & {Total Estimate} &{$\Varone$} & $\Vartwo$ & $\Varthree$ & {$\Varmsmt$} & $\Vtilde$\\ 
&(kt/y) &\multicolumn{4}{c|}{($[\text{kt/y}]^2$)} & ($[\text{kt/y}]^2$) \\ \hline 
Population & 118.981 & 128.069 & 0.950 & 0.040 & 17.502 & 146.561 \\ 
  CO SWB & 0.333 & 0.003 & 0.040 & 0.000 & 0.024 & 0.066 \\ 
  Gas SWB & 0.233 & 0.015 & 0.002 & 0.000 & 0.006 & 0.022 \\ 
  Gas MWB (group) & 5.737 & 2.080 & 0.006 & 0.000 & 0.699 & 2.786 \\ 
  Gas MWB (effluent) & 14.958 & 1.814 & 0.058 & 0.003 & 0.518 & 2.393 \\ 
  Mixed OG Battery & 0.392 & 0.000 & 0.000 & 0.000 & 0.006 & 0.006 \\ 
  Water Hub Battery & 0.216 & 0.028 & 0.000 & 0.000 & 0.040 & 0.068 \\ 
  GP Sweet & 19.275 & 5.718 & 0.061 & 0.001 & 2.615 & 8.395 \\ 
  GP Acid Gas Low & 4.331 & 0.339 & 0.018 & 0.000 & 0.086 & 0.444 \\ 
  Compressor Stations & 46.937 & 102.678 & 0.384 & 0.007 & 9.852 & 112.920 \\ 
  CO MWB & 0.942 & 0.019 & 0.007 & 0.000 & 0.024 & 0.051 \\ 
  CTF and GGS & 0.797 & 0.114 & 0.007 & 0.001 & 0.023 & 0.144 \\ 
  Disposal/storage & 0.456 & 0.067 & 0.001 & 0.000 & 0.008 & 0.076 \\ 
  MS & 0.231 & 0.005 & 0.006 & 0.000 & 0.004 & 0.015 \\ 
  Other GP and LNGP & 9.349 & 1.823 & 0.060 & 0.001 & 0.615 & 2.500 \\ 
  TF and NGL & 0.030 & 0.000 & 0.001 & 0.000 & 0.000 & 0.001 \\ 
  Wells & 14.765 & 13.366 & 0.298 & 0.028 & 2.812 & 16.502 \\ 
   \hline
\end{tabular}
\end{table}

\subsection{Results for Multi-stage IPW with Stage II Uncertainty and No Measurement Error}

\begin{table}[H]
\centering
\caption{Estimates of the total methane emissions for 2021, variance estimates by uncertainty source, and total variance estimates by stratum}
\begin{tabular}{c|c|cccc|c}
\multirow{2}{*}{Stratum} & {Total Estimate} &{$\Vhonee$} & $\Vhtwoo$ & $\Vhthreee$ & {$\Varmsmt$} & $\Vhtott$\\ 
&(kt/y) &\multicolumn{4}{c|}{($[\text{kt/y}]^2$)} & ($[\text{kt/y}]^2$) \\ \hline 
Population & 107.927 & 94.667 & 18.278 & 0.147 & 0.000 & 113.093 \\ 
  CO SWB & 0.180 & 0.001 & 0.001 & 0.001 & 0.000 & 0.003 \\ 
  Gas SWB & 0.225 & 0.012 & 0.003 & 0.000 & 0.000 & 0.015 \\ 
  Gas MWB (group) & 4.745 & 1.025 & 0.353 & 0.001 & 0.000 & 1.378 \\ 
  Gas MWB (effluent) & 14.166 & 1.399 & 1.179 & 0.006 & 0.000 & 2.583 \\ 
  Mixed OG Battery & 0.387 & 0.000 & 0.008 & 0.000 & 0.000 & 0.008 \\ 
  Water Hub Battery & 0.078 & 0.000 & 0.000 & 0.002 & 0.000 & 0.002 \\ 
  GP Sweet & 16.972 & 3.986 & 1.003 & 0.040 & 0.000 & 5.029 \\ 
  GP Acid Gas Low & 3.783 & 0.266 & 0.114 & 0.001 & 0.000 & 0.382 \\ 
  Compressor Stations & 44.753 & 79.815 & 13.505 & 0.007 & 0.000 & 93.327 \\ 
  CO MWB & 0.759 & 0.001 & 0.085 & 0.000 & 0.000 & 0.087 \\ 
  CTF and GGS & 0.776 & 0.105 & 0.007 & 0.001 & 0.000 & 0.114 \\ 
  Disposal/storage & 0.384 & 0.038 & 0.016 & 0.000 & 0.000 & 0.054 \\ 
  MS & 0.158 & 0.003 & 0.000 & 0.000 & 0.000 & 0.004 \\ 
  Other GP and LNGP & 8.459 & 0.402 & 1.544 & 0.003 & 0.000 & 1.948 \\ 
  TF and NGL & 0.025 & 0.000 & 0.001 & 0.000 & 0.000 & 0.001 \\ 
  Wells & 12.078 & 7.614 & 0.460 & 0.085 & 0.000 & 8.159 \\ 
   \hline
\end{tabular}
\end{table}

\subsection{Results for Multi-stage \Hajek\ with Stage II Uncertainty and No Measurement Error}

\begin{table}[H]
\centering
\caption{Estimates of the total methane emissions for 2021, variance estimates by uncertainty source, and total variance estimates by stratum}
\begin{tabular}{c|c|cccc|c}
\multirow{2}{*}{Stratum} & {Total Estimate} &{$\Vhonee$} & $\Vhtwoo$ & $\Vhthreee$ & {$\Varmsmt$} & $\Vhtott$\\ 
&(kt/y) &\multicolumn{4}{c|}{($[\text{kt/y}]^2$)} & ($[\text{kt/y}]^2$) \\ \hline 
Population & 115.860 & 101.050 & 19.585 & 0.003 & 0.000 & 120.638 \\ 
  CO SWB & 0.237 & 0.001 & 0.007 & 0.000 & 0.000 & 0.008 \\ 
  Gas SWB & 0.225 & 0.012 & 0.003 & 0.000 & 0.000 & 0.015 \\ 
  Gas MWB (group) & 5.702 & 1.794 & 0.185 & 0.000 & 0.000 & 1.980 \\ 
  Gas MWB (effluent) & 14.719 & 1.489 & 1.196 & 0.000 & 0.000 & 2.686 \\ 
  Mixed OG Battery & 0.386 & 0.000 & 0.008 & 0.000 & 0.000 & 0.008 \\ 
  Water Hub Battery & 0.078 & 0.001 & 0.000 & 0.000 & 0.000 & 0.002 \\ 
  GP Sweet & 19.180 & 5.130 & 1.294 & 0.000 & 0.000 & 6.423 \\ 
  GP Acid Gas Low & 4.149 & 0.297 & 0.119 & 0.000 & 0.000 & 0.416 \\ 
  Compressor Stations & 46.623 & 82.792 & 14.613 & 0.001 & 0.000 & 97.406 \\ 
  CO MWB & 0.880 & 0.003 & 0.088 & 0.000 & 0.000 & 0.091 \\ 
  CTF and GGS & 0.795 & 0.106 & 0.011 & 0.000 & 0.000 & 0.117 \\ 
  Disposal/storage & 0.450 & 0.056 & 0.017 & 0.000 & 0.000 & 0.073 \\ 
  MS & 0.194 & 0.004 & 0.001 & 0.000 & 0.000 & 0.005 \\ 
  Other GP and LNGP & 9.127 & 0.755 & 1.483 & 0.000 & 0.000 & 2.238 \\ 
  TF and NGL & 0.025 & 0.000 & 0.001 & 0.000 & 0.000 & 0.001 \\ 
  Wells & 13.092 & 8.610 & 0.558 & 0.001 & 0.000 & 9.170 \\ 
   \hline
\end{tabular}
\end{table}

\subsection{Results for Multi-stage IPW with No Stage II Uncertainty and No Measurement Error}

\begin{table}[H]
\centering
\caption{Estimates of the total methane emissions for 2021, variance estimates by uncertainty source, and total variance estimates by stratum}
\begin{tabular}{c|c|cccc|c}
\multirow{2}{*}{Stratum} & {Total Estimate} &{$\Vhonee$} & $\Vhtwoo$ & $\Vhthreee$ & {$\Varmsmt$} & $\Vhtott$\\ 
&(kt/y) &\multicolumn{4}{c|}{($[\text{kt/y}]^2$)} & ($[\text{kt/y}]^2$) \\ \hline 
Population & 107.927 & 107.507 & 0.045 & 0.147 & 0.000 & 107.699 \\ 
  CO SWB & 0.180 & 0.001 & 0.000 & 0.001 & 0.000 & 0.002 \\ 
  Gas SWB & 0.225 & 0.013 & 0.000 & 0.000 & 0.000 & 0.013 \\ 
  Gas MWB (group) & 4.745 & 1.136 & 0.000 & 0.001 & 0.000 & 1.137 \\ 
  Gas MWB (effluent) & 14.166 & 1.626 & 0.000 & 0.006 & 0.000 & 1.632 \\ 
  Mixed OG Battery & 0.387 & 0.000 & 0.000 & 0.000 & 0.000 & 0.000 \\ 
  Water Hub Battery & 0.078 & 0.000 & 0.000 & 0.002 & 0.000 & 0.002 \\ 
  GP Sweet & 16.972 & 4.145 & 0.008 & 0.040 & 0.000 & 4.193 \\ 
  GP Acid Gas Low & 3.783 & 0.286 & 0.000 & 0.001 & 0.000 & 0.287 \\ 
  Compressor Stations & 44.753 & 90.900 & 0.033 & 0.007 & 0.000 & 90.940 \\ 
  CO MWB & 0.759 & 0.014 & 0.000 & 0.000 & 0.000 & 0.015 \\ 
  CTF and GGS & 0.776 & 0.108 & 0.001 & 0.001 & 0.000 & 0.109 \\ 
  Disposal/storage & 0.384 & 0.044 & 0.000 & 0.000 & 0.000 & 0.044 \\ 
  MS & 0.158 & 0.003 & 0.000 & 0.000 & 0.000 & 0.004 \\ 
  Other GP and LNGP & 8.459 & 1.210 & 0.002 & 0.003 & 0.000 & 1.214 \\ 
  TF and NGL & 0.025 & 0.000 & 0.000 & 0.000 & 0.000 & 0.000 \\ 
  Wells & 12.078 & 8.022 & 0.000 & 0.085 & 0.000 & 8.106 \\ 
   \hline
\end{tabular}
\end{table}

\subsection{Results for Multi-stage \Hajek\ with No Stage II Uncertainty and No Measurement Error}

\begin{table}[H]
\centering
\caption{Estimates of the total methane emissions for 2021, variance estimates by uncertainty source, and total variance estimates by stratum}
\begin{tabular}{c|c|cccc|c}
\multirow{2}{*}{Stratum} & {Total Estimate} &{$\Vhonee$} & $\Vhtwoo$ & $\Vhthreee$ & {$\Varmsmt$} & $\Vhtott$\\ 
&(kt/y) &\multicolumn{4}{c|}{($[\text{kt/y}]^2$)} & ($[\text{kt/y}]^2$) \\ \hline  
Population & 115.860 & 114.792 & 0.176 & 0.003 & 0.000 & 114.971 \\ 
  CO SWB & 0.237 & 0.002 & 0.001 & 0.000 & 0.000 & 0.003 \\ 
  Gas SWB & 0.225 & 0.013 & 0.000 & 0.000 & 0.000 & 0.013 \\ 
  Gas MWB (group) & 5.702 & 1.853 & 0.001 & 0.000 & 0.000 & 1.854 \\ 
  Gas MWB (effluent) & 14.719 & 1.719 & 0.006 & 0.000 & 0.000 & 1.725 \\ 
  Mixed OG Battery & 0.386 & 0.000 & 0.000 & 0.000 & 0.000 & 0.000 \\ 
  Water Hub Battery & 0.078 & 0.002 & 0.000 & 0.000 & 0.000 & 0.002 \\ 
  GP Sweet & 19.180 & 5.329 & 0.048 & 0.000 & 0.000 & 5.377 \\ 
  GP Acid Gas Low & 4.149 & 0.318 & 0.002 & 0.000 & 0.000 & 0.319 \\ 
  Compressor Stations & 46.623 & 94.754 & 0.076 & 0.001 & 0.000 & 94.831 \\ 
  CO MWB & 0.880 & 0.016 & 0.001 & 0.000 & 0.000 & 0.017 \\ 
  CTF and GGS & 0.795 & 0.108 & 0.004 & 0.000 & 0.000 & 0.112 \\ 
  Disposal/storage & 0.450 & 0.062 & 0.000 & 0.000 & 0.000 & 0.063 \\ 
  MS & 0.194 & 0.004 & 0.000 & 0.000 & 0.000 & 0.004 \\ 
  Other GP and LNGP & 9.127 & 1.529 & 0.005 & 0.000 & 0.000 & 1.534 \\ 
  TF and NGL & 0.025 & 0.000 & 0.000 & 0.000 & 0.000 & 0.000 \\ 
  Wells & 13.092 & 9.085 & 0.031 & 0.001 & 0.000 & 9.117 \\ 
   \hline
\end{tabular}
\end{table}

\bibliography{jrss-refs}